\documentclass[useAMS,usenatbib,referee]{mn2e}

\usepackage{amsmath, amssymb}
\usepackage[dvips]{graphicx}
\usepackage{epsfig}
\usepackage{color}

\newcommand{\rot}{\bmath{\nabla} \times}

\newcommand{\rlight}{r_{\rm L}}
\newcommand{\Rs}{R_{\rm s}}
\newcommand{\as}{a_{\rm s}}

\newcommand{\ephi}{\bmath{e}_\varphi}

\newcommand{\aap}{A\&A}
\newcommand{\mnras}{MNRAS}

\newcommand{\apj}{ApJ}
\newcommand{\apjl}{ApJL}

\newcommand{\pasj}{PASJ}

\newcommand{\prd}{Physical Review D}

\newcommand{\jcap}{JCAP}
\newcommand{\sovast}{Soviet. Astr.}

\title[Pulsar magnetospheres]{General-relativistic force-free pulsar magnetospheres} 
\author[J. P\'etri]{J.  P\'etri$^{1}$
\thanks{E-mail: jerome.petri@astro.unistra.fr} \\
  $^{1}$Observatoire astronomique de Strasbourg, Universit\'e de Strasbourg, CNRS, UMR 7550, 11 rue de l'universit\'e, F-67000 Strasbourg, France.}

\begin{document}

\date{Accepted . Received ; in original form }

\pagerange{\pageref{firstpage}--\pageref{lastpage}} 
\pubyear{2015}

\maketitle

\label{firstpage}

\begin{abstract}
Pulsar magnetospheres are shaped by ultra-relativistic electron/positron plasmas flowing in a strong magnetic field and subject to strong gravitational fields. The former induces magnetospheric currents and space charges responsible for the distortion of the electromagnetic field based on pure electrodynamics. The latter induces other perturbations in these fields based on space-time curvature. The force-free approximation describes the response of this magnetosphere to the presence of currents and charges and has been investigated by many authors. In this context, general relativity has been less discussed to quantify its influence on the neutron star electrodynamics. It is the purpose of this paper to compute general-relativistic force-free pulsar magnetospheres for realistic magnetic field configurations such as the inclined dipole. We performed time-dependent simulations of Maxwell equations in the 3+1 formalism of a stationary background metric in the slow-rotation approximation. We computed the resulting Poynting flux depending on the ratio~$R/\rlight$ and on frame-dragging through the spin parameter~$\as$, $R$ is the neutron star radius and $\rlight$ the light-cylinder radius. Both effects act together to increase the total Poynting flux seen by a distant observer by a factor up to~2 depending on the rotation rate. Moreover we retrieve the $\sin^2\chi$ dependence of this luminosity, $\chi$ being the obliquity of the pulsar, as well as a braking index close to $n=3$. We also show that the angular dependence of the Poynting flux scales like $\sin^2\vartheta$ for the aligned rotator but like $\sin^4\vartheta$ for the orthogonal rotator, $\vartheta$ being the colatitude.
\end{abstract}


\section{Introduction}

Neutron stars are places where strongly magnetized and relativistic plasmas are embedded in a strong gravitational field. It is usually admitted that electron-positron pairs fill their magnetosphere.  Both plasma and gravity significantly impact on the structure of this magnetosphere. Curvature and frame-dragging effects are indeed important due to the high compactness of neutron stars. For typical models of neutron star interiors, the compactness is about $\Xi = \Rs/R \approx 0.5$ where $\Rs=2\,G\,M/c^2$ is the Schwarzschild radius, $M$ is the mass of the neutron star, $R$ its radius, $G$ the gravitational constant and $c$ the speed of light. From these estimates, we expect a departure from the flat space-time results as large as 15-20\%.

Although we are not confronted to the problem of an event horizon as for black hole electrodynamics, frame-dragging effects and curvature of space will distort the magnetic field in the surrounding of the neutron star. As a first step towards a realistic description of neutron star magnetospheres, the force-free approximation is often invoked as a drastic simplification of the problem in the presence of a strongly magnetized plasma. Recently, several groups performed numerical simulations of pulsar magnetospheres in this limit without gravitational effects \citep{2006ApJ...648L..51S, 2006MNRAS.368L..30M, 2006MNRAS.367...19K, 2012MNRAS.420.2793K, 2012MNRAS.424..605P, 2012MNRAS.423.1416P}.

Our attention now also focuses on the general-relativistic extension of this force-free picture. Indeed, using a pseudo-spectral discontinuous Galerkin approach, \cite{2015MNRAS.447.3170P} computed general-relativistic solutions to the force-free monopole and split monopole fields. \cite{PhysRevD.89.084045} performed the first general-relativistic simulations matching the interior solution (within the neutron star) to the exterior solution (the force-free magnetosphere). In a similar way, \cite{2015MNRAS.447.2821P} investigated general-relativistic equilibria of neutron stars including twisted magnetic fields by solving a Grad-Shafranov equation. A description of pulsar magnetospheres in terms of Grad-Shafranov equation was already proposed by \cite{2005MNRAS.358..998K} in general relativity in a way similar to black holes in the MHD limit. \cite{2009MNRAS.395..443A} and \cite{2010MNRAS.408..490M} computed the oscillations of the magnetosphere of an aligned rotator in general relativity. Moreover \cite{2008ApJ...684.1359M} studied the influence of a monopole on the acceleration of particles in the polar caps extending previous work by \cite{1990SvAL...16..286B} and \cite{1990SvA....34..133M}. \cite{2014MNRAS.445.2500G} used tools from differential geometry such as the exterior calculus to deduce some general properties of force-free magnetospheres of black holes and neutron stars. Finally MHD simulations of neutron star magnetospheres were performed by \cite{2006MNRAS.367...19K} and later by \cite{2013MNRAS.435L...1T}. But these force-free or MHD approaches only represent dissipationless systems not allowing an interchange of energy between the plasma and the electromagnetic field. Therefore a new degree of freedom is required in the description of the magnetosphere. The resistivity of the plasma could account for such dissipation. It was indeed included in the FFE description by \cite{2008JCAP...11..002G}, \cite{2012ApJ...749....2K}, \cite{2012ApJ...746...60L} and even in MHD by \cite{2013MNRAS.431.1853P}. 

The latest improvements have been made using particle in cell (PIC) simulations. As the simplest example, \cite{2007MNRAS.376.1460W,2011MNRAS.418..612W} focused on the plasma electrostatic response to the magnetospheric structure in the aligned dipole rotator using a specially designed electronic chip to efficiently compute the coulomb interactions between charged particles. Special attention to the Y-point was given by \cite{2010PASJ...62..131U}. With this PIC approach, 
\cite{2009ApJ...690...13M} retrieved the quiet configuration depicted by the electrosphere found by \cite{1985MNRAS.213P..43K} and \cite{2002A&A...384..414P}. On the other hand, full PIC simulations made by \cite{2015MNRAS.449.2759B} show that up to 20\% of the spin-down luminosity can be carried by the particle depending on the particle injection rate. Transition from an electrosphere to the force-free magnetosphere is observed depending on the injection process \citep{2014ApJ...785L..33P}. \cite{2014ApJ...795L..22C} found the same conclusions by investigation of the aligned magnetosphere including $e^\pm$ discharge and radiation. According to \cite{2015ApJ...801L..19P}, the oblique rotator in PIC simulations also shows a $\sin^2\chi$ dependence as in the force-free limit. \cite{2014MNRAS.441.1879P} also made a comparison between force-free and MHD simulations of the magnetosphere. They were led to the same conclusions. Moreover, they self-consistently computed the torque exerted on the neutron star surface from their simulations. \cite{2012PASJ...64...43Y} investigated more deeply polar cap, slot gap, and outer gap models to explain pulsed emission. All the above magnetospheres contain a current sheet wobbling around the equatorial plane, see however \cite{2014PASJ...66...25T} for an alternative approach to pulsar magnetosphere without this current sheet.

Our goal in this paper is to quantify precisely the distortion induced by general-relativistic effects, namely curvature of space-time and frame dragging in the force-free limit. To this end, we solve the time-dependent Maxwell equations in curved space-time in spherical coordinates for a background dipolar magnetic field anchored into the neutron star. Consequently, we use the 3+1~formalism of electrodynamics. The code algorithm is reminded in depth in Section~\ref{sec:Algorithme}. The code is then tested in 1D against known analytical solutions in Cartesian coordinates in Section~\ref{sec:Tests}. Application of the code to vacuum and force-free regimes are presented respectively in Section~\ref{sec:Vacuum} and Section~\ref{sec:FFE}. Conclusions and ongoing works are drawn in Section~\ref{sec:Conclusion}.

\section{Numerical algorithm}
\label{sec:Algorithme}

Our code is based on a multi-domain decomposition technique employing discontinuous Galerkin methods as exposed in \cite{2015MNRAS.447.3170P}. This high-order finite volume method was successfully applied to the monopole and split monopole field in the force-free regime. Unknown variables were computed in an orthonormal basis. In our current version of the code, we improved its flexibility by allowing a freedom in the choice of the metric. To this end, we use covariant and contravariant components instead of the physical components to express Maxwell equations, as exposed in the next paragraph. Therefore we are able to compute any kind of electromagnetic structure in a prescribed background curved space in three dimensions once the spatial metric $\gamma_{ab}$, the lapse function~$\alpha$ and the shift vector~$\bmath \beta$ have been set up. Remember that in our convention, indexes with letters from $a$ to $h$ span only the absolute space coordinates whereas letters starting from $i$ span the four dimensional spacetime. Details about the implementation of the spatial discretization, the temporal integration, the initial and boundary conditions and the filtering process are given below for completeness.

\subsection{Maxwell equations in component form}

In order to treat on a same foot any kind of curvilinear coordinate system with the same code, we write Maxwell equations in component form adapted to an absolute space with spatial coordinates~$x^a$ and a time coordinate~$t$ as described by an observer with four velocity~$n^i$. The time evolution of the electric and magnetic fields, $\mathbf D$ and $\mathbf B$, is therefore
\begin{subequations}
 \begin{align}
  \partial_t ( \sqrt{\gamma} \, D^a ) & = \varepsilon^{abc} \, \partial_b H_c - \sqrt{\gamma} \, J^a \\
  \partial_t ( \sqrt{\gamma} \, B^a ) & = - \varepsilon^{abc} \, \partial_b E_c
 \end{align}
\end{subequations}
with the constitutive relations
\begin{subequations}
 \begin{align}
  \varepsilon_0 \, E_a & = \alpha \, D_a + \varepsilon_0 \, c \, \sqrt{\gamma} \, \varepsilon_{abc} \, \beta^b \, B^c \\
  \mu_0 \, H_a & = \alpha \, B_a - \sqrt{\gamma} \, \varepsilon_{abc} \, \beta^b \, D^c / (\varepsilon_0 \, c)
 \end{align}
\end{subequations}
and the constraint equations
\begin{subequations}
 \begin{align}
  \frac{1}{\sqrt{\gamma}} \, \partial_a ( \sqrt{\gamma} \, D^a ) & = \rho \\
  \frac{1}{\sqrt{\gamma}} \, \partial_a ( \sqrt{\gamma} \, B^a ) & = 0 .
 \end{align}
\end{subequations}
$\varepsilon_{abc}$ and $\varepsilon^{abc}$ are the fully antisymmetric three dimensional Levi-Civita tensors. $\varepsilon_0$ and $\mu_0$ are respectively the vacuum permittivity and permeability. The source terms are the charge density~$\rho$ and the current density~$J^a$.
The spacetime geometry is described by the lapse function~$\alpha$, the shift vector~$\bmath\beta$ and the spatial metric~$\gamma_{ab}$, with the definition $\gamma = {\rm det}(\gamma_{ab})$. In the slow rotation approximation frequently used for neutron stars, the metric is essentially described by two parameters: the Schwarzschild radius defined by
\begin{equation}
 \Rs = \frac{2\,G\,M}{c^2}
\end{equation}
and the spin parameter~$\as$. A reasonable choice for spherically symmetric and homogeneous neutron stars is
\begin{equation}
\label{eq:SpinParameter}
 \frac{\as}{\Rs} = \frac{2}{5} \, \frac{R}{\Rs} \, \frac{R}{\rlight}
\end{equation}
where $\rlight=c/\Omega$ is the light cylinder radius and $\Omega$ the neutron star rotation rate.
The spatial metric is given in spherical Boyer-Lindquist coordinates by
\begin{equation}
  \label{eq:Metric3D}
  \gamma_{ab} =
  \begin{pmatrix}
    \alpha^{-2} & 0 & 0 \\
    0 & r^2 & 0 \\
    0 & 0 & r^2 \sin^2\vartheta
  \end{pmatrix}
\end{equation}
the lapse function by
\begin{equation}
  \label{eq:Lapse}
  \alpha = \sqrt{ 1 - \frac{\Rs}{r} }
\end{equation}
and the shift vector by
\begin{subequations}
 \begin{align}
  \label{eq:Shift}
  c \, \bbeta = & - \omega \, r \, \sin\vartheta \, \ephi \\
  \omega = & \frac{\as\,\Rs\,c}{r^3}
 \end{align}
\end{subequations}
In contravariant components the only non vanishing term of the shift vector is $\beta^\varphi = - \omega/c$. From observations we have $R \leq 0.1 \, \rlight$ and realistic equations of state for supranuclear matter gives $\Rs \approx 0.5\,R$. In that case the upper limit for the spin parameter would be $\as \leq 0.08 \, \Rs \lesssim 0.1 \, \Rs$. In the remainder of this paper, the spin parameter $\as$ is constrained by the rotation speed of the neutron star through eq.~(\ref{eq:SpinParameter}). In all the subsequent simulations about neutron star magnetospheres, we will distinguish between two sets of run, the first one concerning electrodynamics in Newtonian gravity with $\Rs=\as=0$ and the second one including general relativity with a Schwarzschild radius fixed to half the stellar radius~$R$, $\Rs=R/2$, and the spin parameter $\as$ according to eq.~(\ref{eq:SpinParameter}).

We now remind the basic features of our pseudo-spectral discontinuous Galerkin finite element method including the expansion on to vector spherical harmonics for divergencelessness fields, the exact imposition of boundary conditions on the neutron star surface and outgoing waves at the outer boundary, the explicit time stepping with a third order strong stability preserving Runge-Kutta (SSPRK) integration scheme, the spectral filtering in the longitudinal and latitudinal directions and the limiter applied in the radial direction. The radial part is solved with a high-order finite volume scheme whereas the spherical part is solved through a pseudo-spectral approach.

\subsection{One dimensional scalar conservation law}

We briefly remind the basic feature of a discontinuous Galerkin approach by focusing on a one dimension non-linear scalar conservation law. Consider therefore a scalar field denoted by~$u$ with a physical flux function denoted by~$f(u)$ such that the conservation of~$u$ is expressed as a partial differential equation written as
\begin{equation}
\label{eq:Conservation}
 \partial_t u + \partial_x f(u) = 0.
\end{equation}
This equation has to be solved for any time $t\geq0$ and for all $x\in[a,b]$ where $[a,b]$ is the computational domain. Note that in our code $x$ should be interpreted as the radial coordinate~$r$. We subdivide the domain~$[a,b]$ in $K$~cells not necessarily of the same length. In each of these cells which we denote by~$D^k$ with $k\in[0..K-1]$, the solution is expanded on to a basis of spatial functions~$\phi^k_i$ such that the approximate solution in the cell~$k$ reads
\begin{equation}
 u^k(x,t) = \sum_{i=0}^{N_p} u^k_i(t) \, \phi^k_i(x)
\end{equation}
valid in the cell~$k$ given by the interval~$[x^k_l,x^k_r]$. The basis possesses $N_p+1$ functions. The spatial method is therefore of order~$N_p+1$. After injecting this expansion into the conservation law equation~(\ref{eq:Conservation}) and projecting on to the basis functions~$\phi_i^k$, performing two successive integrations by part in each cell independently and introducing a numerical flux $f^*$ we arrive at the semi-discrete system to be solved in matrix notation
\begin{equation}
 \label{eq:GalerkinDiscontinueMat}
 \mathcal{M}^k \, \partial_t \mathcal{U}^k + \mathcal{S}^k \, \mathcal{F}^k = [ ( f - f^*) \, \mathcal{P}^k ]_{x^k_l}^{x^k_r}
\end{equation}
with the following matrices
\begin{subequations}
\begin{align}
 \mathcal{M}_{ij}^k & = \int_{x^k_l}^{x^k_r} \phi_i^k \, \phi_j^k \, dx \\
 \mathcal{S}_{ij}^k & = \int_{x^k_l}^{x^k_r} \phi_i^k \, \partial_x \phi_j^k \, dx
\end{align}
\end{subequations}
that can be computed analytically and exactly and with $\mathcal{P}^k$ the column vector of basis polynomials. Note that for an orthogonal basis, the mass matrix~$\mathcal{M}^k$ is diagonal hence very easy to invert. Inverting the mass matrix~$\mathcal{M}^k$, each coefficient of the expansion evolves according to the first order ordinary differential equation
\begin{equation}
 \label{eq:GalerkinDiscontinueMatricielle}
 \partial_t \mathcal{U}^k + (\mathcal{M}^k)^{-1} \, \mathcal{S}^k \, \mathcal{F}^k = (\mathcal{M}^k)^{-1} \, [ ( f - f^*) \, \mathcal{P}^k ]_{x^k_l}^{x^k_r} .
\end{equation}
The state of the art in discontinuous Galerkin methods resides in the choice of the numerical flux~$f^*$ which has to satisfy several stability and consistency properties.

\subsection{The grid}

The arbitrary nature of the radial coordinate is used to fix small volumes close to the neutron star whereas larger shells are sufficient farther away. To be more specific, we employ the usual Fourier transform in the $\{\vartheta,\varphi\}$ directions and expand the radial coordinate into $K$ sub-intervals, the boundary of each cell is given by~$[r_g^k,r_d^k]$ with $k\in[0..K-1]$ dividing the global interval $[R_1,R_2]$ into non necessarily equal sub-intervals. Let us assume that the computational domain is comprised between the neutron star surface at~$R_1=R$ and an arbitrary outer radius~$R_2>R_1$. The spherical shell is decomposed into $K$ cells but with increasing thickness. We introduce two temporary variables $y_1=\log(R_1/\rlight)$ and $y_2=\log(R_2/\rlight)$ and a logarithmic thickness by $h=(y_2-y_1)/K$. Each cell, labelled with a superscript~$k$, possesses then two interfaces located at
\begin{subequations}
 \begin{align}
  r_g^k & = e^{y_1+k\,h} \\
  r_d^k & = e^{y_1+(k+1)\,h} .
 \end{align}
\end{subequations}
The thickness of the cell labelled~$k$ is $h^k=r_d^k-r_g^k$.
In that way, the ratio between the size of two successive cells is constant and equal to $e^h$.

\subsection{Vector expansion and divergencelessness constraint on $\bmath{B}$}

We use again an expansion of the vector fields $\bmath{B}$ and $\bmath{D}$. Indeed, electric and magnetic fields are expanded onto vector spherical harmonics (VSH) according to
\begin{subequations}
 \begin{align}
   \label{eq:D_vhs}
  \bmath{D} & = \sum_{l=0}^\infty\sum_{m=-l}^l \left(D^r_{lm} \, \bmath{Y}_{lm} + D^{(1)}_{lm} \, \bmath{\Psi}_{lm} + D^{(2)}_{lm} \, \bmath{\Phi}_{lm}\right) \\
  \label{eq:B_vhs}
  \bmath{B} & = \sum_{l=0}^\infty\sum_{m=-l}^l \left(B^r_{lm} \, \bmath{Y}_{lm} + B^{(1)}_{lm} \, \bmath{\Psi}_{lm} + B^{(2)}_{lm} \, \bmath{\Phi}_{lm}\right)
\end{align}
\end{subequations}
Such expansion is done in each cell. However, in order to deal with the divergencelessness of the magnetic field whatever the configuration of the electromagnetic field, loaded or not with plasma it is more appropriate to use an expansion of $\bmath{B}$ into
\begin{equation}
  \label{eq:B_div0}
  \bmath{B} = \sum_{l=1}^\infty\sum_{m=-l}^l \rot [f^B_{lm}(r,t) \, \bmath{\Phi}_{lm}] + g^B_{lm}(r,t) \, \bmath{\Phi}_{lm} 
\end{equation}
where $\{f^B_{lm}(r,t), g^B_{lm}(r,t)\}$ are the expansion coefficients of $\bmath{B}$. See \cite{2013MNRAS.433..986P} for more details about vector spherical harmonics.

\subsection{Numerical flux}

As in any other finite volume scheme, communication between cells goes through a numerical flux~$f^*$ chosen to resolve as accurately as possibly the conservation laws. However, for non-linear problems, exact solutions to the associated Riemann problem are usually difficult to solve and computationally expensive. In order to be as general as possible, we decided to use the simple but robust Lax-Friedrich flux such that
\begin{equation}
 f^* = \frac{1}{2} \, [ f(u_d^{k-1}) + f(u_g^k) - C \, (u_g^k-u_d^{k-1}) ]
\end{equation}
with the constant $C=max_{u}|f'(u)|$ interpreted as the maximum speed for the waves. Fortunately, for high-order methods, the choice of the numerical flux does not impact drastically on the results. Actually, the simulations become insensitive to the exact choice of the flux for high-order methods.

\subsection{Slope Limiter}

The slope limiting technique is adapted from the classical finite volume community. The idea is to reduce spurious oscillations that arise from the non-linear evolution or from sharp discontinuities in the solution. The most basic total variation diminishing (TVD) limiters are usually too dissipative for higher-order schemes. \cite{Toro2009} detailed several TVD schemes with application to simple problems. We refer the reader to this book for more information about TVD methods. Another less stringent technique uses a total variation bound (TVB) method \citep{1989JCoPh..84...90C}. The latter does not guaranty strict cancellation of oscillations but only weaken them whereas the former completely avoids oscillations but at the cost of reducing to a low-order scheme. Unfortunately TVB methods introduce one more parameter, often depicted by the capital letter~$M$. Moreover the value of this parameter is very problem dependent, related to the second spatial derivative of the solution, therefore a priori unknown. Eventually we tried another limiting procedure of higher order and called moment limiter, as described by \cite{2007JCoPh.226..879K}. This technique successively limits the coefficients of the expansions in polynomials from the highest orders down to the lowest coefficients. The limiting is aborted as soon as a coefficient remains unchanged.

\subsection{Filtering}

The limiter cannot be applied in the latitudinal and longitudinal direction simply because there is no domain decomposition in those directions. We use the classical spherical harmonic expansion. The force-free problem being non-linear due to the electric current in the source terms, we expect the solution to develop sharp gradients or discontinuities also in the spherical directions. It is therefore compulsory to get rid of these high frequencies by some filtering procedure. This is achieved by adding a small damping factor to the high order coefficients of the expansion in $Y_{lm}$. Filtering is performed at each time step. We use an exponential filter in the directions~$\{\vartheta,\varphi\}$ given by the general expression
\begin{equation}
  \label{eq:Filtre}
  \sigma(\eta) = \textrm{e}^{-\alpha\,\eta^\beta}
\end{equation}
where the variable~$\eta$ ranges between 0 and 1. For instance, in the latitudinal direction $\eta = l/(N_{\vartheta}-1)$ for $l\in[0..N_{\vartheta}-1]$, $l$ being the index of the coefficient $c_{lm}$ in the spherical harmonic expansion $f(\vartheta,\varphi) = \sum_{l,m=0}^{N_{\vartheta}-1,N_{\varphi}-1} c_{lm} \,Y_{lm}(\vartheta,\varphi)$ and $\{N_{\vartheta},N_\varphi\}$ the number of collocation points in the spherical direction (latitude and longitude). The parameter $\alpha$ (not to be confused with the lapse function) is adjusted to values not too large in order to avoid errors in the solution but also not too small in order to sufficiently damp oscillations. 

The above mentioned exponential filter of order~$\beta$ does not strictly satisfy the condition for the smoothing factors as explained in \cite{Canuto2006}. However, for numerical purposes we choose $\alpha$ such that $\textrm{e}^{-\alpha}$ is numerically zero i.e. below the machine accuracy $\varepsilon$. In practice, we choose $\alpha=36$ assuming double precision computation with $\epsilon\approx10^{-15}$. The order~$\beta$ of the smoothing influences the dissipation rate in the solution. The low order multipole components are weakly damped and correspond to large scale structures. If the solution shows fine scale structures, the filtering has to be minimized.

\subsection{Time integration}

One of the strength of pseudo-spectral methods is that they replace a set of partial differential equations (PDE) by a larger set of ordinary differential equations (ODE) for the unknown collocation points or spectral coefficients. Schematically, it can be written  as
\begin{equation}
  \label{eq:ODE}
  \frac{d\bmath{u}}{dt} = f(t,\bmath{u})
\end{equation}
with appropriate initial and boundary conditions. $\bmath{u}$ represents the vector of unknown functions either evaluated at the collocation points or the spectral coefficients. We use a third order strong stability preserving Runge-Kutta scheme advancing the unknown functions~$\bmath{u}$ in time. See \cite{Hesthaven2008} for more details about these time integration schemes especially including the popular second and third order symbolically written as SSPRK2 and SSPRK3.

\subsection{Boundary conditions}

As in \cite{2014MNRAS.439.1071P} we put exact boundary conditions on the star. In general relativity the correct jump conditions at the stellar surface, continuity of the normal component of the magnetic field~$B^{r}$ and continuity of the tangential component of the electric field~$\{D^{\vartheta}, D^{\varphi}\}$ are such that
\begin{subequations}
  \label{eq:CLimites}
\begin{align}
  B^{r}(t,R,\vartheta,\varphi) & = B^{r}_0(t,\vartheta,\varphi) \\
  D^{\vartheta}(t,R,\vartheta,\varphi) & = - \varepsilon_0 \, \frac{\Omega-\omega}{\alpha^2} \, \sin\vartheta \, B^{r}_0(t,\vartheta,\varphi) \\
  D^{\varphi}(t,R,\vartheta,\varphi) & = 0 .
\end{align}
\end{subequations}
Note the slight difference in these expressions with respect to our previous papers because here we use tensor components in a non orthonormal basis.
The continuity of $B^{r}$ automatically implies the correct boundary treatment of the electric field. $B^{r}_0(t,\vartheta,\varphi)$ represents the, possibly time-dependent, radial magnetic field imposed by the star, let it be monopole, split monopole, oblique dipole or multipole.

Maxwell equations, explicitly written, are
\begin{subequations}
 \begin{align}
  \partial_t ( \sqrt{\gamma} \, D^r ) & = \partial_\vartheta H_\varphi - \partial_\varphi H_\vartheta - \sqrt{\gamma} \, J^r \\
  \partial_t ( \sqrt{\gamma} \, D^\vartheta ) & = \partial_\varphi H_r - \partial_r H_\varphi - \sqrt{\gamma} \, J^\vartheta \\
  \partial_t ( \sqrt{\gamma} \, D^\varphi ) & = \partial_r H_\vartheta - \partial_\vartheta H_r - \sqrt{\gamma} \, J^\varphi \\
  \partial_t ( \sqrt{\gamma} \, B^r ) & = \partial_\varphi E_\vartheta - \partial_\vartheta E_\varphi \\
  \partial_t ( \sqrt{\gamma} \, B^\vartheta ) & = \partial_r E_\varphi - \partial_\varphi E_r \\
  \partial_t ( \sqrt{\gamma} \, B^\varphi ) & = \partial_\vartheta E_r - \partial_r E_\vartheta .
 \end{align}
\end{subequations}
We look for the characteristics propagating along the radial direction. To this end, we isolate expressions containing the radial propagation that is $\partial_r$ and $\partial_t$. Eliminating all useless terms for this radial propagation, the system reduces to
\begin{subequations}
 \begin{align}
  \partial_t ( \sqrt{\gamma} \, D^\vartheta ) + \partial_r H_\varphi & = 0 \\
  \partial_t ( \sqrt{\gamma} \, D^\varphi ) - \partial_r H_\vartheta & = 0 \\
  \partial_t ( \sqrt{\gamma} \, B^\vartheta ) - \partial_r E_\varphi & = 0 \\
  \partial_t ( \sqrt{\gamma} \, B^\varphi ) + \partial_r E_\vartheta & = 0 .
 \end{align}
\end{subequations}
The covariant components of the spatial vectors $ \bmath D$ and $\bmath B$ are giving by lowering the indexes such that
\begin{subequations}
 \begin{align}
 D_\vartheta & = \gamma_{\vartheta r} \, D^r + \gamma_{\vartheta \vartheta} \, D^\vartheta + \gamma_{\vartheta \varphi} \, D^\varphi \\
 D_\varphi & = \gamma_{\varphi r} \, D^r + \gamma_{\varphi \vartheta} \, D^\vartheta + \gamma_{\varphi \varphi} \, D^\varphi \\
 B_\vartheta & = \gamma_{\vartheta r} \, B^r + \gamma_{\vartheta \vartheta} \, B^\vartheta + \gamma_{\vartheta \varphi} \, B^\varphi \\
 B_\varphi & = \gamma_{\varphi r} \, B^r + \gamma_{\varphi \vartheta} \, B^\vartheta + \gamma_{\varphi \varphi} \, B^\varphi .
 \end{align}
\end{subequations}
Injecting the constitutive relations into the evolution equations, we find
\begin{subequations}
 \begin{align}
  \partial_t ( \sqrt{\gamma} \, \mu_0 \, D^\vartheta ) + \partial_r ( \alpha \, B_\varphi ) - \partial_r \left( \frac{\sqrt{\gamma}}{\varepsilon_0\,c} \, (\beta^r\,D^\vartheta - \beta^\vartheta \, D^r) \right) & = 0 \\
  \partial_t ( \sqrt{\gamma} \, \mu_0 \, D^\varphi ) - \partial_r ( \alpha \, B_\vartheta ) +  \partial_r \left( \frac{\sqrt{\gamma}}{\varepsilon_0\,c} \, (\beta^\varphi\,D^r - \beta^r \, D^\varphi) \right) & = 0 \\
  \partial_t ( \sqrt{\gamma} \, \varepsilon_0 \, B^\vartheta ) - \partial_r ( \alpha \, D_\varphi ) - \partial_r ( \varepsilon_0 \, c \, \sqrt{\gamma} \, (\beta^r\,B^\vartheta - \beta^\vartheta \, B^r)) & = 0 \\
  \partial_t ( \sqrt{\gamma} \, \varepsilon_0 \, B^\varphi ) + \partial_r ( \alpha \, D_\vartheta ) + \partial_r ( \varepsilon_0 \, c \, \sqrt{\gamma} \, (\beta^\varphi\,B^r - \beta^r \, B^\varphi) ) & = 0 .
 \end{align}
\end{subequations}
Defining the unknown vector
\begin{equation}
\bmath{U} =
 \begin{pmatrix}
  \sqrt{\gamma} \, \mu_0 \, D^\vartheta \\
  \sqrt{\gamma} \, \mu_0 \, D^\varphi \\
  \sqrt{\gamma} \, \varepsilon_0 \, B^\vartheta \\
  \sqrt{\gamma} \, \varepsilon_0 \, B^\varphi
 \end{pmatrix}
\end{equation}
the system can be cast into the conservative form $\partial_t \bmath{U} + \partial_r (A\,\bmath{U}) = 0$ with
\begin{equation}
A =
 \begin{pmatrix}
  -c\,\beta^r & 0 & \alpha\,\gamma_{\varphi\vartheta} / (\varepsilon_0\,\sqrt{\gamma}) &  \alpha\,\gamma_{\varphi\varphi} / (\varepsilon_0\,\sqrt{\gamma}) \\
  0 & -c\,\beta^r & -\alpha\,\gamma_{\vartheta\vartheta} / (\varepsilon_0\,\sqrt{\gamma}) &  - \alpha\,\gamma_{\varphi\vartheta} / (\varepsilon_0\,\sqrt{\gamma}) \\
  - \alpha\,\gamma_{\varphi\vartheta} / (\mu_0\,\sqrt{\gamma}) & - \alpha\,\gamma_{\varphi\varphi} / (\mu_0\,\sqrt{\gamma}) & -c\,\beta^r & 0 \\
  \alpha\,\gamma_{\vartheta\vartheta} / (\mu_0\,\sqrt{\gamma}) & \alpha\,\gamma_{\vartheta\varphi} / (\mu_0\,\sqrt{\gamma}) & 0 & -c\,\beta^r
 \end{pmatrix}.
\end{equation}
We restrict ourselves to metrics with vanishing components~$\gamma_{\vartheta\varphi}$. Therefore, the eigenvalues for the electromagnetic waves propagating in vacuum in general relativity are given by
\begin{equation}
 ( - \beta^r \pm \alpha\,\sqrt{\frac{\gamma_{\vartheta\vartheta} \, \gamma_{\varphi\varphi}}{\gamma}} ) \, c
\end{equation}
and the eigenvectors by
\begin{subequations}
 \begin{align}
  ( \pm \sqrt{\frac{\gamma_{\varphi\varphi} \, \mu_0}{\gamma_{\vartheta\vartheta} \, \varepsilon_0}}, 0, 0, 1) & \ \ \ ( 0, \pm \sqrt{\frac{\gamma_{\vartheta\vartheta} \, \mu_0}{\gamma_{\varphi\varphi} \, \varepsilon_0}}, 1, 0) .
 \end{align}
\end{subequations}
For the slowly rotating metric in spherical Boyer-Lindquist coordinates these expressions simplify for the wave speed
\begin{equation}
  ( - \beta^r \pm \alpha^2 ) \, c
\end{equation}
and for the conserved quantities propagating along the characteristics
\begin{subequations}
 \begin{align}
  \varepsilon_0 \, c\, \sin\vartheta \, B^\varphi \pm D^\vartheta & \\
  \varepsilon_0 \, c\, B^\vartheta \pm \sin\vartheta \, D^\varphi & .
 \end{align}
\end{subequations}
The outer boundary condition cannot be handled exactly. We need to make some approximate assumptions about the outgoing waves we want to enforce in order to prevent reflections from this artificial outer boundary. Using the Characteristic Compatibility Method (CCM) described in \cite{Canuto2007} and neglecting the frame-dragging effect far from the neutron star, the radially propagating characteristics are given to good accuracy by their flat spacetime expressions as
\begin{eqnarray}
  \label{eq:CCM1}
  D^{\vartheta} \pm \varepsilon_0 \, c\, \sin\vartheta \, B^{\varphi} & ; & \sin\vartheta \, D^{\varphi} \pm \varepsilon_0 \, c\, B^{\vartheta}.
\end{eqnarray}
In order to forbid ingoing waves we ensure that
\begin{subequations}
\begin{align}
  \label{eq:CCM2}
  D^{\vartheta} - \varepsilon_0 \, c\, \sin\vartheta \, B^{\varphi} & = 0 \\
  \label{eq:CCM3}
  \sin\vartheta \, D^{\varphi} + \varepsilon_0 \, c\, B^{\vartheta} & = 0
\end{align}
whereas the other two characteristics are found by
\begin{align}
  \label{eq:CCM4}
  D^{\vartheta} + \varepsilon_0 \, c\, \sin\vartheta \, B^{\varphi} & = D^{\vartheta}_{\rm PDE} + \varepsilon_0 \, c\, \sin\vartheta \, B^{\varphi}_{\rm PDE} \\
  \label{eq:CCM5}
  \sin\vartheta \, D^{\varphi} - \varepsilon_0 \, c\, B^{\vartheta} & = \sin\vartheta \, D^{\varphi}_{\rm PDE} - \varepsilon_0 \, c\, B^{\vartheta}_{\rm PDE}
\end{align}
\end{subequations}
the subscript $_{\rm PDE}$ denoting the values of the electromagnetic field obtained by straightforward time advancing without care of any boundary condition. The new corrected values are deduced from the solution of the linear system made of equations~(\ref{eq:CCM2})-(\ref{eq:CCM5}).

\subsection{Initial conditions}

The rotation of the neutron star is switched on smoothly as in our previous works, see \cite{2012MNRAS.424..605P, 2014MNRAS.439.1071P, 2015MNRAS.447.3170P}. Its spin frequency increases slowly starting from zero. Taking an evolution of the spin frequency as
\begin{equation}
 \Omega(t) = 
\begin{cases}
\sin^2\left(\frac{t}{8}\right) \text{ for } t\leq4\,\upi \\ 
1  \text{ for } t\geq4\,\upi
\end{cases}
\end{equation}
starting at zero speed avoids the initial discontinuity in the electric field. The normalized period at full rotation speed is $2\,\upi$. The spin frequency as well as its first derivative are smooth at the initial time of the simulation~$t=0$. The final time of all simulation runs is set to $T_{\rm f}=12\,\upi$ and corresponds to five full rotations of the neutrons star. This is sufficient for the system to settle down to its quasi-stationary state.

\section{Tests}
\label{sec:Tests}

In our update of the discontinuous Galerkin code, we use a tensorial (covariant) formalism to compute the vector components in order to specify freely the background metric. It is therefore possible to describe any kind of coordinate system through this metric. In particular, we can test our code in a one dimensional Cartesian coordinate system and compare our simulation results with exact analytical solutions in the force-free limit. To this aim, several special cases are usually checked as reported in \cite{2002MNRAS.336..759K,2004MNRAS.350..427K} and in \cite{2013PhRvD..88j4031P}.

We performed our tests starting with the initial conditions given in these aforementioned works. For completeness, we reproduce these test cases in the subsequent paragraphs. Results are shown in Fig.~\ref{fig:Test_1D_FFE1} and Fig.~\ref{fig:Test_1D_FFE2}. The simulation points are indicated with blue crosses whereas the analytical solutions, if they exist, are shown in red solid line. For simulation purposes, we use normalized units for the electromagnetic field by setting $c = \mu_0 = \varepsilon_0 = 1$. The background metric is Minkowski space-time in Cartesian coordinates. Let us describe shortly each case. From the analysis of force-free electrodynamics, as reported by \cite{2002MNRAS.336..759K}, we expect only two kind of waves: fast waves that are similar to electromagnetic waves in vacuum and Alfven waves similar to those of relativistic MHD.
We do not follow strictly the initial conditions prescribed in the above literature, instead of a linear or sinusoidal variation when required, we preferred a tangent hyperbolic spatial dependence.

\subsection{Alfven wave}

Alfven waves are important solutions to force-free electrodynamics. In the frame of the wave itself, the initial conditions are given by
\begin{subequations}
 \begin{align}
 \phi & = 1.15 + 0.15 \, \tanh (10\,x) \\
  E'_x & = - c \, B'_z = - \phi \\
  E'_y & = 0 \\
  E'_z & = B'_x = B'_y = 1 .
 \end{align}
\end{subequations}
These expressions need to be boosted into the observer frame via a Lorentz transform. The wave therefore propagates to the left at a speed given in the observer frame by 0.5. The simulation results are shown in fig.~\ref{fig:Test_1D_FFE1}, upper panel. The wave speed is accurately caught by our algorithm.

\subsection{Current sheet}

\begin{figure*}
 \centering
 \input{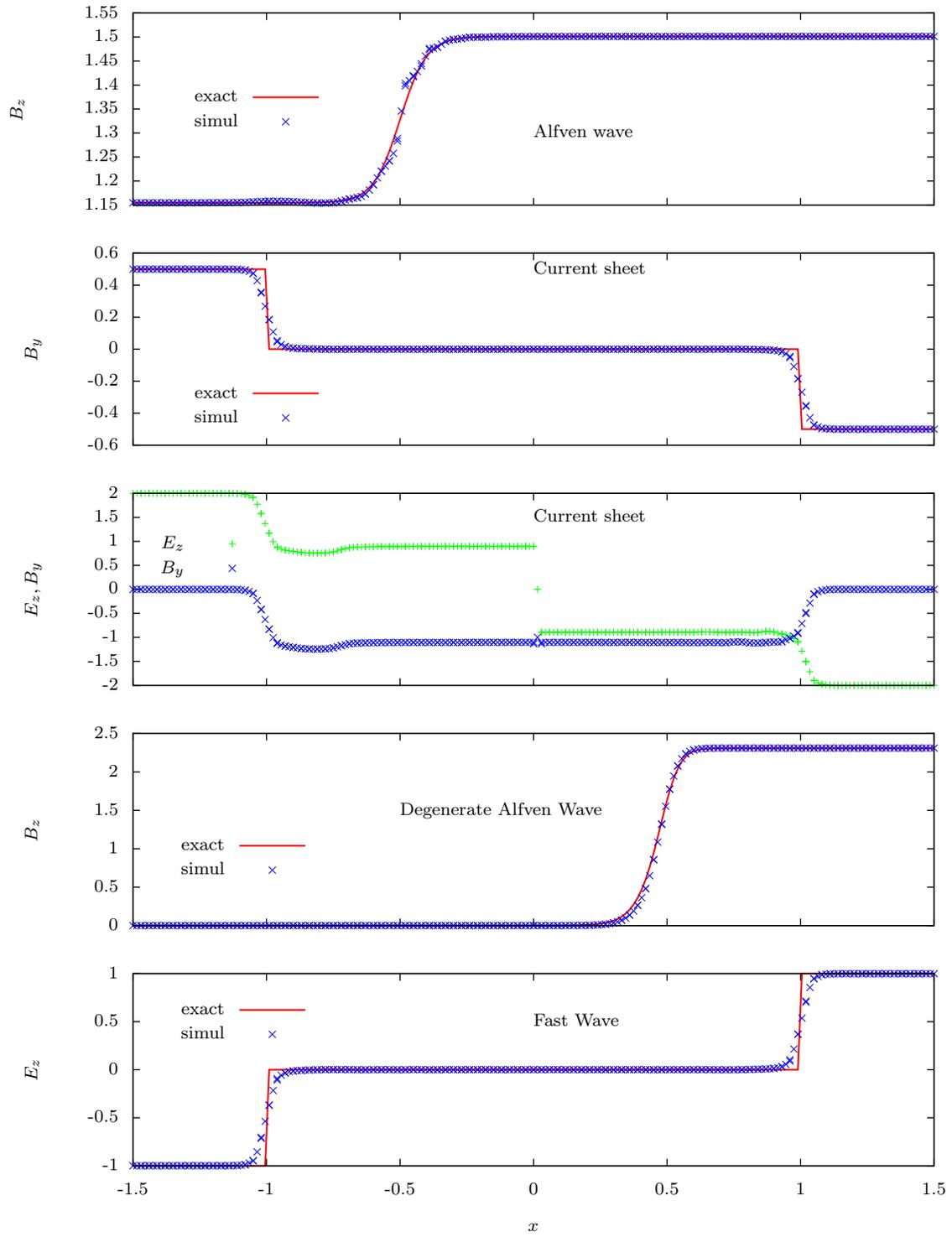}
 \caption{Numerical test of the code for several 1D Cartesian problems, first set.}
 \label{fig:Test_1D_FFE1}
\end{figure*}

A current sheet corresponds to a solution with an initial discontinuity in the magnetic field. As a starting point, we use the following setup
\begin{subequations}
 \begin{align}
  B_y & =
  \begin{cases}
   B_0 & \textrm{ if } x\leqslant1 \\
   -B_0 & \textrm{ if } x\geqslant1
  \end{cases} \\
  E_x & = E_y = E_z = B_z = 0 \\
  B_x & = 1 .
 \end{align}
\end{subequations}
The temporal evolution will then depend on the value of $B_0$. In a first run we used $B_0=0.5$. This launches two fast waves moving in opposite directions at the speed of light, see second panel from top of fig.~\ref{fig:Test_1D_FFE1}. In a second run we used $B_0=2$. Two fast waves are still allowed but force-free electrodynamics has no solution, see middle panel of fig.~\ref{fig:Test_1D_FFE1}.

\subsection{Degenerate Alfven wave}

A degenerate Alfven wave can be produced in the following manner. In the frame of the wave itself, the initial conditions are given by
\begin{subequations}
 \begin{align}
 \phi & = \upi \, \frac{1+\tanh (10\,x)}{4} \\
  E'_x & = E'_y = E'_z = 0 = c \, B'_x \\
  B'_y & = 2 \, \cos\phi \\
  B'_z & = 2 \, \sin\phi .
 \end{align}
\end{subequations}
These expressions need to be boosted into the observer frame via a Lorentz transform. The wave therefore propagates to the right at a speed given in the observer frame by 0.5, see second panel from the bottom of fig.~\ref{fig:Test_1D_FFE1}.

\subsection{Fast wave}

This test corresponds to two fast waves given initially by a discontinuity at $x=0$ such that
\begin{subequations}
 \begin{align}
  E_z & = - c\, B_y =
  \begin{cases}
   1 & \textrm{ if } x\leqslant1 \\
   -1 & \textrm{ if } x\geqslant1
  \end{cases} \\
  E_x & = E_y = B_z = 0 \\
  B_x & = 1 .
 \end{align}
\end{subequations}
One propagates to the right and the other to the left, both at the speed of light.
We have chosen a discontinuity to show the effect of filtering on a discontinuous solution. The smoothing is clearly apparent in the lower panel of fig.~\ref{fig:Test_1D_FFE1} but the speed of the wave is exactly reproduced.

\subsection{Stationary Alfven wave}

An Alfven wave with zero speed is given by
\begin{subequations}
 \begin{align}
 \phi & = 1.25 + 0.25 \, \tanh (10\,x) \\
  E_x & = - c\, B_y = - \phi \\
  E_y & = c\,B_x = 1 \\
  E_z & = 0.0 \\
  B_z & = 2 .
 \end{align}
\end{subequations}
It corresponds to a stationary Alfven state remaining in our example around $x=0$ as seen in the upper panel of fig.~\ref{fig:Test_1D_FFE2}.

\subsection{Stationary state}

This test corresponds to a stationary state given by
\begin{subequations}
 \begin{align}
  E_x & = c\, B_z = \tanh (10\,x) \\
  E_y & = E_z = B_x = 0 \\
  B_y & = 1 .
 \end{align}
\end{subequations}
No time evolution is expected. The only effect could be a smearing due to numerical dissipation caused by our filtering procedure. Actually, inspecting the second panel from top of fig.~\ref{fig:Test_1D_FFE2}, the stationary state is maintained to very good accuracy for a long time.

\subsection{Three waves}

It is possible to get two fast waves propagating in opposite direction and a stationary Alfven wave in the same run for the special initial conditions given by
\begin{subequations}
 \begin{align}
  \mathbf{B} =
  \begin{cases}
   (1,1.5,3.5) & \textrm{ if } x\leqslant0 \\
   (1,3,3) & \textrm{ if } x>0
  \end{cases} \\
  \mathbf{E} =
  \begin{cases}
   (-1,-0.5,0.5) & \textrm{ if } x\leqslant0 \\
   (-1.5,2,-1.5) & \textrm{ if } x>0
  \end{cases}
 \end{align}
\end{subequations}
This magnetic configuration splits into two fast discontinuities propagating in opposite directions at the speed of light and a stationary Alfven wave staying at $x=0$, second panel from bottom of fig.~\ref{fig:Test_1D_FFE2}.

\subsection{FFE breakdown}

\begin{figure*}
 \centering
 \input{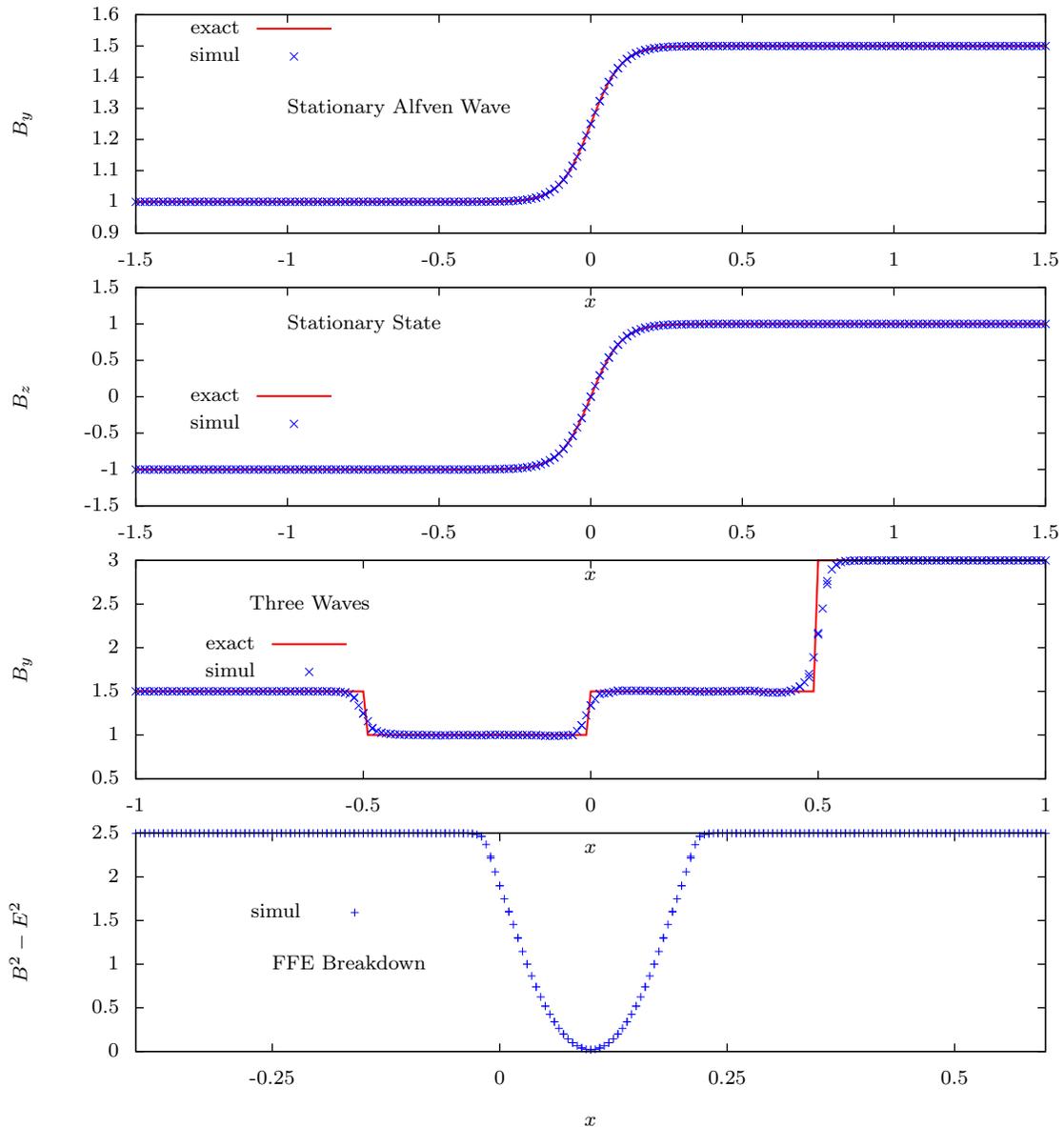}
 \caption{Numerical test of the code for several 1D Cartesian problems, second set.}
 \label{fig:Test_1D_FFE2}
\end{figure*}

In some cases, the FFE approximation can be violated during the evolution process of the plasma. For instance, with the initial conditions given by 
\begin{subequations}
 \begin{align}
  \mathbf{B} & =
  \begin{cases}
   (1,1,1) & \textrm{ if } x \leqslant 0 \\
   (1,1-10\,x,1-10\,x) & \textrm{ if } 0 \geqslant x \leqslant 0.2 \\
   (1,-1,-1) & \textrm{ if } x \geqslant 0.2
  \end{cases} \\
  \mathbf{E} & = (0,0.5,-0.5) .
 \end{align}
\end{subequations}
after a short time of about $\Delta t=0.02$, the requirement $E^2<B^2$ everywhere becomes difficult to be fulfilled. $B^2-E^2$ starts to vanish around $x=0.1$, bottom panel of fig.~\ref{fig:Test_1D_FFE2}.

Going through all the above tests, we showed that the code reproduces the expected analytical solutions to good accuracy. We switch now to the 3D spherical case concerning neutron star electromagnetic fields.

\section{Vacuum magnetospheres}
\label{sec:Vacuum}

Before going to the most interesting case of a force-free pulsar magnetosphere, we check again our algorithm against known vacuum solutions, given exactly in flat space-time and to high accuracy in curved space-time. In all the simulations shown below, the Poynting flux $\bmath{S}$ is used as an efficient and simple diagnostic tool to check the convergence of the code. For comparison between different magnetic topologies as well as for the discrepancies between Newtonian (N) and general-relativistic (GR) cases, this luminosity is always computed through a sphere~$\mathcal{S}_{\rm L}$ of radius equal to the light-cylinder radius thus
\begin{equation}
 L = \int_{\mathcal{S}_{\rm L}} (\bmath E \wedge \bmath H )^{\hat r} \, dS
\end{equation}
where the hat $\hat r$ means the physical component of the Poynting vector and $dS=r^2\,d\Omega$ the infinitesimal surface element on the sphere~$\mathcal{S}_{\rm L}$.

\subsection{Monopole}

\begin{figure}
\begin{center}
\input{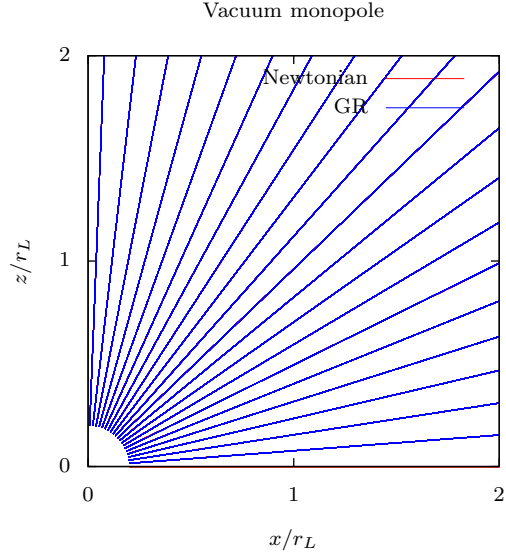}
\end{center}
\caption{Meridional field lines for the vacuum monopole in flat space-time, red line, and in the slow rotation approximation, blue line, with $R/\rlight=0.2$. The Newtonian case and the relativistic case overlap perfectly.}
\label{fig:Vacuum_Monopole1}
\end{figure}

The monopole field represents the archetypal solution for which an exact analytical expression exists in flat space-time, in vacuum but most importantly for the force-free rotator. It serves as a good test of efficiency and accuracy of any algorithm. Thus we started with a monopole field in vacuum. The results found by our numerical code agree well with the analytical expressions. As an illustration, we show some magnetic field lines in the meridional plane for the Newtonian case $\Rs=0$ and the slow rotation approximation with $\Rs/R=0.5$ in fig.~\ref{fig:Vacuum_Monopole1} for a rotator possessing a rotational speed in normalized units given by~$R/\rlight=0.2$. The radial structure of the field topology is not influenced by general relativity. The field lines remain straight as in flat space-time. Indeed, the radial component of the magnetic field is the only non vanishing part and is simply expressed by 
\begin{equation}
 B^r = \alpha \, B \, \frac{R^2}{r^2}
\end{equation}
where $B$ is the magnetic field strength as measured by a distant observer. The Poynting flux vanishes after a transition phase during which the electric field builds up to its final stage.

\subsection{Dipole}

\begin{figure}
\begin{center}
\input{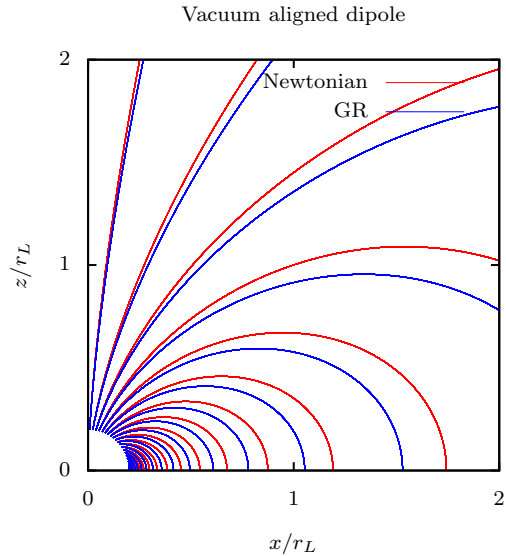}
\end{center}
\caption{Meridional field lines for the vacuum aligned dipole in flat space-time, red line, and in the slow rotation approximation, blue line, with $R/\rlight=0.2$. The location of the foot of each line is the same in both runs.}
\label{fig:Vacuum_Dipole_Aligned1}
\end{figure}

The lowest order magnetic field topology anchored into a neutron star is dipolar. Thus we investigate the case of an aligned dipole in vacuum. Magnetic field lines in the meridional plane for the Newtonian case and the slow rotation approximation are depicted in fig.~\ref{fig:Vacuum_Dipole_Aligned1} for $R/\rlight=0.2$. Compared to flat space-time, general relativity compresses the field lines toward the equatorial plane. The Poynting flux here also vanishes due to the axisymmetry of the configuration.

\begin{figure}
\begin{center}
\input{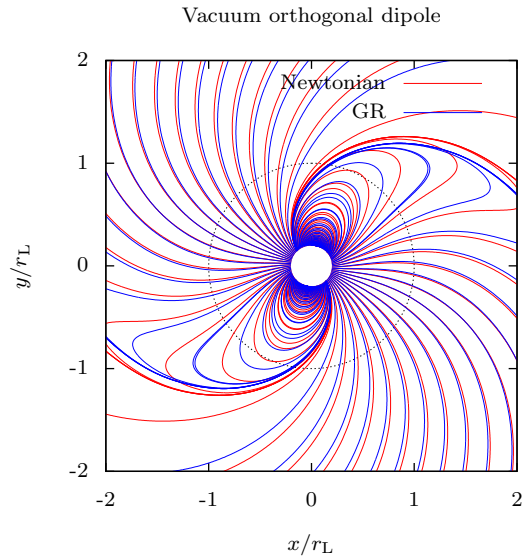}
\end{center}
\caption{Equatorial field lines for the vacuum perpendicular dipole in flat space-time, red line, and in the slow rotation approximation, blue line, with $R/\rlight=0.2$.}
\label{fig:Vacuum_Dipole_Perp1}
\end{figure}

\begin{figure}
\begin{center}
\input{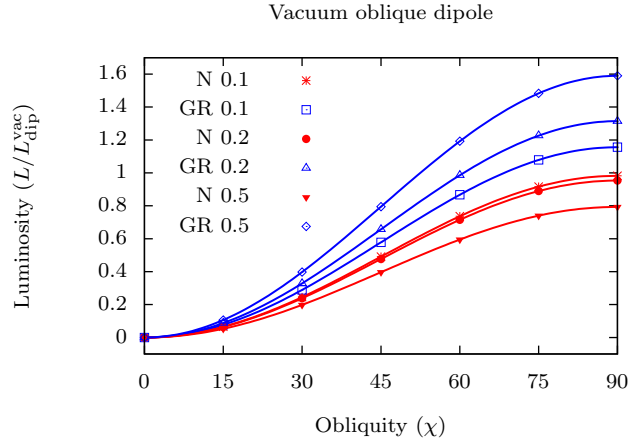}
\end{center}
\caption{Poynting flux of the vacuum oblique rotator for different obliquities~$\chi$ and normalized to $L_{\rm dip}^{\rm vac}$ with $R/\rlight=\{0.1, 0.2, 0.5\}$. Points are taken from the simulations and the solid lines are the best fits, Newtonian in red (N) and general-relativistic in blue (GR).}
\label{fig:Poynting_Dipole_Oblique_Vacuum_R1}
\end{figure}

The oblique rotator is the most interesting case for neutron star magnetosphere. It represents the lowest order multipole anchored in the star although higher order multipoles could exist, see \cite{2015MNRAS.450..714P}. As a special case, we investigate the perpendicular rotator. In order to get an idea of the magnetic field topology, field lines contained in the equatorial plane are shown in fig.~\ref{fig:Vacuum_Dipole_Perp1}, comparing again the Newtonian dipole and the slow rotation dipole with $R/\rlight=0.2$.
The typical spin-down luminosity, used for normalization, is given by the orthogonal vacuum rotator
\begin{equation}
\label{eq:SpinDownDipole}
L_{\rm dip}^{\rm vac} = \frac{8\,\upi}{3} \, \frac{\Omega^4\,B^2\,R^6}{\mu_0\,c^3}.
\end{equation}
To conclude the vacuum simulations, we performed a set of simulations for oblique rotators in the case $R/\rlight=\{0.1, 0.2, 0.5\}$ and for $\chi = \{0\degr,15\degr,30\degr,45\degr,60\degr,75\degr,90\degr\}$ to check the dependence of the luminosity on the geometry. The results are reported in fig.~\ref{fig:Poynting_Dipole_Oblique_Vacuum_R1} for the Newtonian, red symbols, and general-relativistic, blue symbols, gravitational fields. The points are taken from the simulations whereas the solid curves are best fits obtained by adjusting to a $\sin^2\chi$ dependence such that
\begin{equation}
 \frac{L}{L_{\rm dip}^{\rm vac}} = b \, \sin^2\chi .
\end{equation} 
The precise values are reported in Table~\ref{tab:FitFluxVacuum} comparing Newtonian and general-relativistic gravity. As the ratio~$R/\rlight$ decreases, both kind of curves, Newtonian and general-relativistic, tend to the function~$\sin^2\chi$, the former from below and the latter from above.
\begin{table}
\centering
\begin{center}
\begin{tabular}{ccc}
\hline
$R/\rlight$ & Newtonian & GR \\
\hline
\hline
0.1 & 0.982 & 1.156 \\
0.2 & 0.955 & 1.316 \\
0.5 & 0.793 & 1.590 \\
\hline
\end{tabular}
\end{center}
\caption{Best fit parameter~$b$ for the Poynting flux $L(\chi)/L_{\rm dip}^{\rm vac} = b\,\sin^2\chi$ of the vacuum dipole rotator in Newtonian and general-relativistic case.}
\label{tab:FitFluxVacuum}
\end{table}

\begin{figure}
\begin{center}
\input{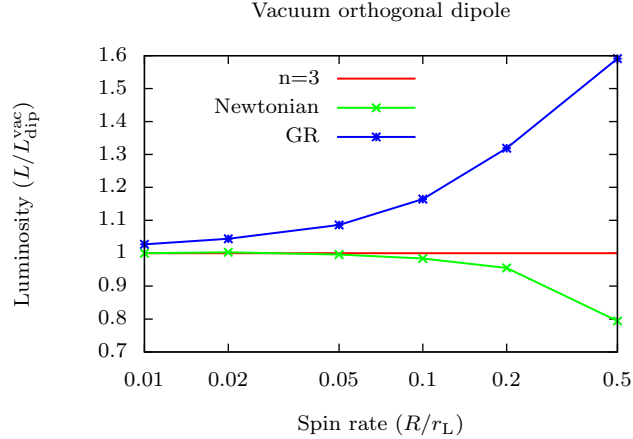}
\end{center}
\caption{Poynting flux of the vacuum orthogonal rotator normalized to $L_{\rm dip}^{\rm vac}$. The different approximations are shown in green for Newtonian and blue for GR and compared to a braking index of $n=3$ in red.}
\label{fig:Poynting_Dipole_Vacuum_Deutsch}
\end{figure}

\begin{figure}
\begin{center}
\input{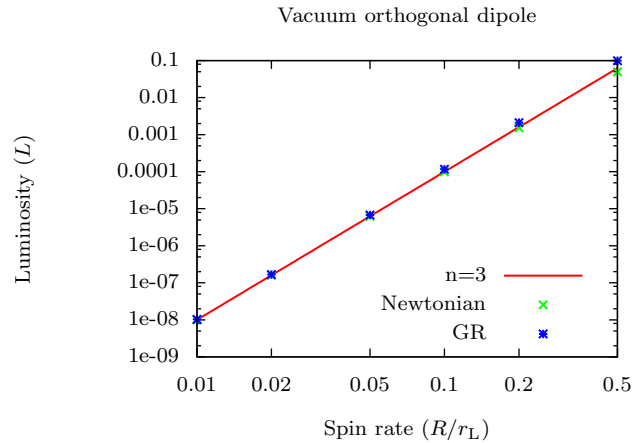}
\end{center}
\caption{Absolute Poynting flux of the vacuum orthogonal dipole rotator. The different approximations are marked by symbols, green for Newtonian and blue for GR and compared to the point dipole expectations for which~$n=3$ in red.}
\label{fig:Indice_Dipole_Vacuum_Deutsch}
\end{figure}

\begin{figure}
\begin{center}
\input{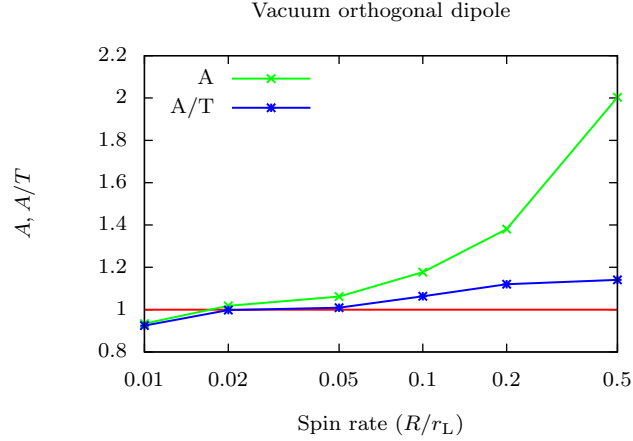}
\end{center}
\caption{Comparison of the increase in total luminosity as obtained by the simulation through the amplification factor~$A=L_{\rm GR}/L_{\rm Newt}$, green solid curve, and as predicted by the increase in the transverse magnetic field~$A/T = (L_{\rm GR}/L_{\rm Newt})/\textrm{theory}$, blue solid curve. For reference the line $L=1$ is also shown in red.}
\label{fig:Dipole_Vacuum_Perp_Explication}
\end{figure}

As a final comparison, we plot the Poynting fluxes expected from the Deutsch solution and the simulations in fig.~\ref{fig:Poynting_Dipole_Vacuum_Deutsch}. We are able to go down to rotation rates as low as $R/\rlight=0.01$ allowing us to span an appreciable range of rotation periods.
The Poynting flux of the vacuum orthogonal rotator is reported in fig.~\ref{fig:Poynting_Dipole_Vacuum_Deutsch} for periods between $R/\rlight=0.01$ and $0.5$. Adjusting the absolute Poynting in fig.~\ref{fig:Indice_Dipole_Vacuum_Deutsch} with a power we can estimate the braking index. Indeed, the associated estimated braking indexes for the Deutsch solution and the simulation are respectively $n=3.05$ and $n=2.93$. Consequently, the point dipole braking index in vacuum, known to be equal to $n=3$ is retrieved to good accuracy in the general-relativistic case.
The increase in total luminosity can in part be explained by the increase in the transverse magnetic field in the vicinity of the light-cylinder. Indeed, in fig.~\ref{fig:Dipole_Vacuum_Perp_Explication} we compute the ratio $L_{\rm GR}/L_{\rm Newt}$ as a solid red curve. We observe an increase by a factor two for the fastest rotator with $R/\rlight=0.5$. Meanwhile, we compared this increase of luminosity to the increase in the physical components of the transverse magnetic field components $\{B^{\hat\vartheta},B^{\hat\varphi}\}$ at the light-cylinder as predicted by the general-relativistic vacuum dipole solution, see the green curve in fig.~\ref{fig:Dipole_Vacuum_Perp_Explication}. Within 10-15\% the increase in luminosity is explained by the strength of the magnetic field at the light-cylinder. Note that we do not take into account possible frame-dragging effects that are stronger for the fastest rotators, explaining the discrepancies for the highest $R/\rlight$.

\begin{figure}
\begin{center}
\input{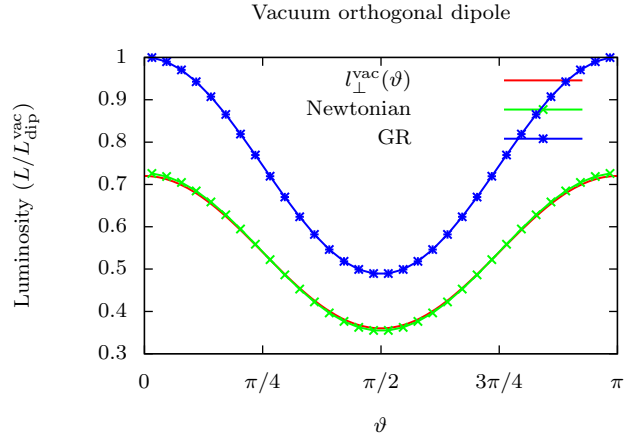}
\end{center}
\caption{Angular dependence of the Poynting flux for the vacuum perpendicular dipole in flat space-time and in the slow rotation approximation with $R/\rlight=0.2$. It corresponds to the flux through the sphere of radius~$5\,\rlight$.}
\label{fig:Vacuum_Dipole_Perp_Flux_Angulaire}
\end{figure}

The angular dependence of the Poynting flux normalized to the total luminosity for a point dipole is given for any inclination~$\chi>0$ at large distance $r\gg\rlight$ by
\begin{equation}
l_{\perp}^{\rm vac}(\vartheta) = \frac{L(\vartheta)}{L_{\rm dip}^{\rm vac}} = \frac{3}{8} \, ( 1 + \cos^2\vartheta) \, \left[ 1 - \left(\frac{R}{\rlight}\right)^2 \right]
\end{equation}
Note that it is independent of the inclination angle~$\chi$ and takes into account corrections in power of $R/\rlight$. For the vacuum perpendicular dipole, in fig.~\ref{fig:Vacuum_Dipole_Perp_Flux_Angulaire}, this analytical expression is shown in solid red line and compared to the Newtonian simulations, green crosses, and to its general-relativistic extension, blue stars. This Poynting flux is extracted from the sphere of radius $r=5\,\rlight$. All the angular dependences of the luminosity shown in this paper are given at this particular radius.

\section{Force-free magnetospheres}
\label{sec:FFE}

The most interesting results concern the plasma screening effect in the pulsar magnetosphere. This is investigated in the force-free approximation as a starting point. Full general relativity is account for by the 3+1 formalism already employed in previous works. We revisit the general-relativistic monopole and split monopole cases before diving into the dipole rotator for arbitrary inclination.

\subsection{Monopole}

\begin{figure}
\begin{center}
\input{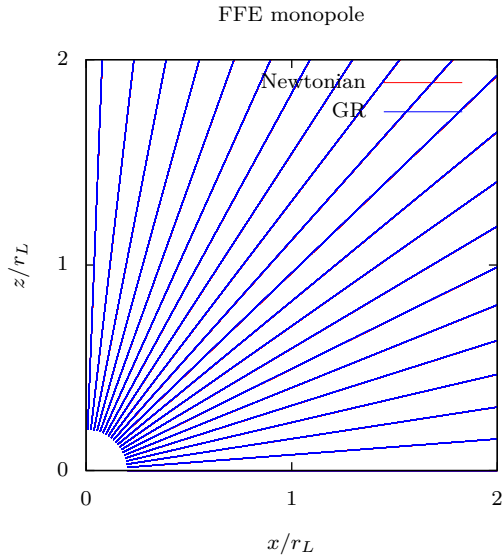}
\end{center}
\caption{Meridional field lines for the force-free monopole in flat space-time, red curve, and in the slow rotation approximation, blue curve. The period of the neutron star is such that $R/\rlight=0.2$. Field lines overlap exactly.}
\label{fig:FFE_Monopole1}
\end{figure}

\begin{figure}
 \centering
 \begin{center}
\input{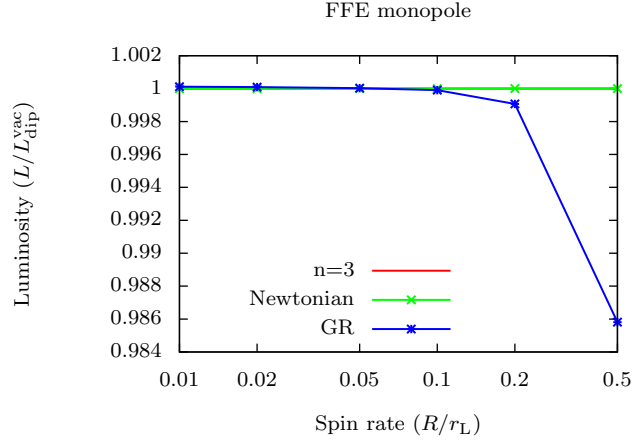}
\end{center}
\caption{Luminosity of the force-free monopole rotator in flat space-time and in the slow rotation approximation with $R/\rlight=\{0.01, 0.02, 0.05, 0.1,0.2, 0.5\}$ and normalized to $L_{\rm dip}^{\rm vac}$. The different approximations are given in the legend and compared to the point dipole expectations $n=3$, red line.}
 \label{fig:Luminosite_FFE_Monopole}
\end{figure}

\begin{figure}
\begin{center}
\input{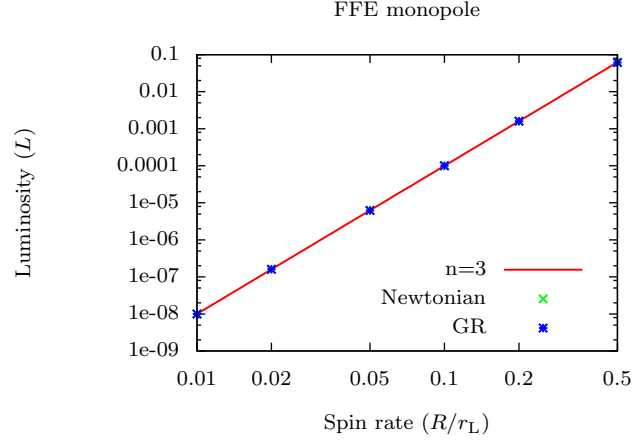}
\end{center}
\caption{Absolute Poynting flux of the force-free monopole. The different approximations are green crosses for Newtonian and blue stars for GR and compared to the point dipole expectations $n=3$ in red.}
\label{fig:Indice_Monopole_FFE}
\end{figure}

\begin{figure}
\begin{center}
\input{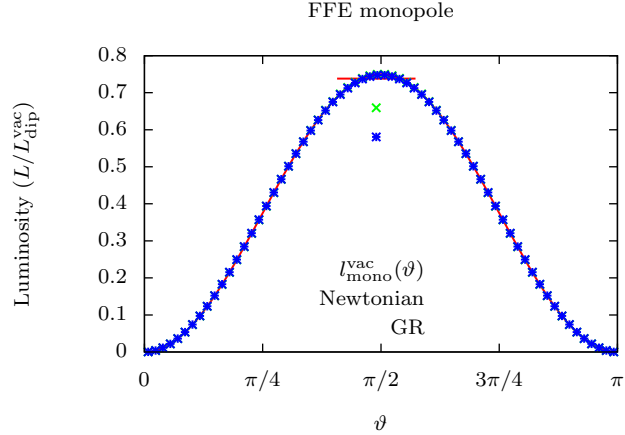}
\end{center}
\caption{Angular dependence of the Poynting flux for the force-free monopole in flat space-time, green crosses, and in the slow rotation approximation, blue stars, with $R/\rlight=0.2$. The fit to $\sin^2\vartheta$ is shown in red.}
\label{fig:FFE_Monopole_Flux_Angulaire}
\end{figure}

\begin{figure}
 \centering
\input{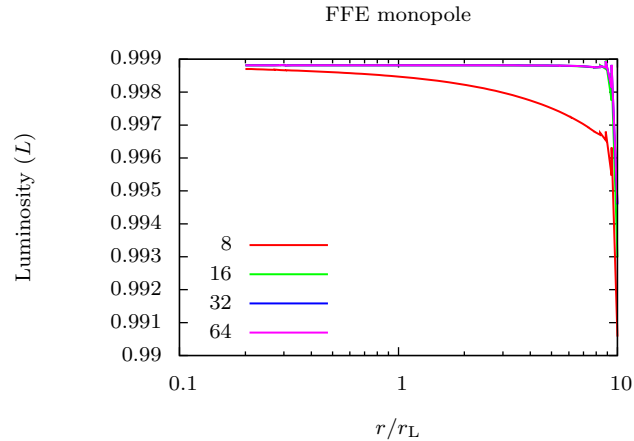}
 \caption{Luminosity of the force-free monopole in the slow rotation approximation with $R/\rlight=0.2$ and for different numerical resolutions $N_\vartheta=\{8,16,32,64\}$ as given in the legend.}
 \label{fig:Luminosite_FFE_Monopole_Dissipation}
\end{figure}

The monopole field is simply described by a radial component decreasing like an inverse square law. This expression is valid in flat as well as in curved space-time. In Boyer-Lindquist coordinates, there is no distinction between both topologies. Indeed, inspecting fig.~\ref{fig:FFE_Monopole1}, the field lines in the meridional plane overlap whatever the rotation rate~$R/\rlight$.
Nevertheless, the Poynting flux, as measured by a distant observer, decreases slightly with increasing spin rate as seen in fig.~\ref{fig:Luminosite_FFE_Monopole}.
To finish the discussion with this set of simulations, the braking indexes for the Newtonian and general-relativistic monopole solutions are compared and equal respectively to $n=3.000$ and $n=2.996$, obtained from fitting the absolute Poynting flux shown in fig.~\ref{fig:Indice_Monopole_FFE} to a power law in~$R/\rlight$.
The angular dependence of the Poynting flux for the force-free monopole normalized to its total luminosity, which is equivalent to the perpendicular vacuum dipole rotator luminosity, is equal to
\begin{equation}
l_{\rm mono}^{\rm vac}(\vartheta) = \frac{L(\vartheta)}{L_{\rm dip}^{\rm vac}} = \frac{3}{4} \, \sin^2\vartheta
\end{equation}
and represented in fig.~\ref{fig:FFE_Monopole_Flux_Angulaire}. The analytical expression in solid red line overlaps with the Newtonian simulations depicted by green crosses and also with the general-relativistic simulations in blue stars.
In order to investigate the influence of the numerical resolution on the accuracy of the simulations, we plot the Poynting flux for the force-free monopole in fig.\ref{fig:Luminosite_FFE_Monopole_Dissipation} for different resolutions by increasing the number of colatitudinal points $N_\vartheta$ in the general-relativistic case and for $R/\rlight=0.2$. Because the solution does not contain any discontinuity, already a low resolution gives satisfactory results. The error is less than 1\%.

\subsection{Split monopole}

\begin{figure}
\begin{center}
\input{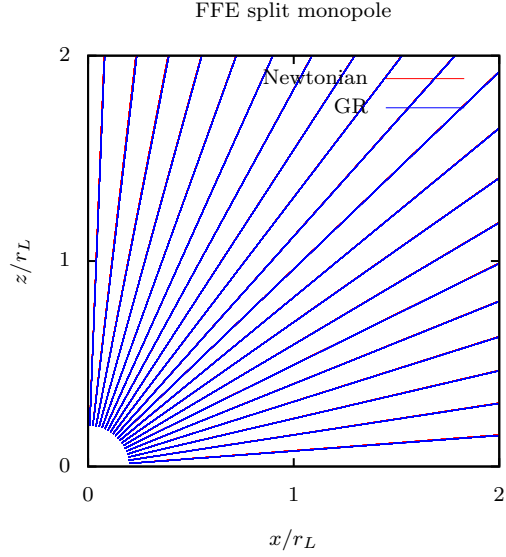}
\end{center}
\caption{Meridional field lines for the force-free split monopole in flat space-time, red line, and in the slow rotation approximation, blue line, with $R/\rlight=0.2$.}
\label{fig:FFE_Split1}
\end{figure}

\begin{figure}
 \centering
\input{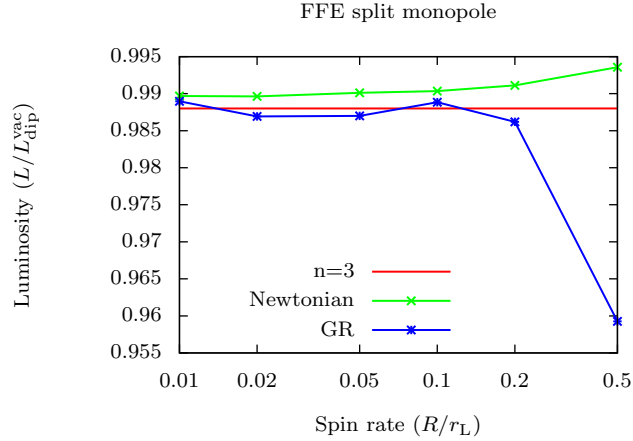}
 \caption{Luminosity of the force-free split monopole in flat space-time, red line, and in the slow rotation approximation, blue line, with $R/\rlight=\{0.01, 0.02, 0.05, 0.1, 0.2, 0.5\}$. The different approximations are shown in green for Newtonian and blue  in GR and compared to the point dipole expectations $n=3$ in red.}
 \label{fig:Luminosite_FFE_Split}
\end{figure}

\begin{figure}
\begin{center}
\input{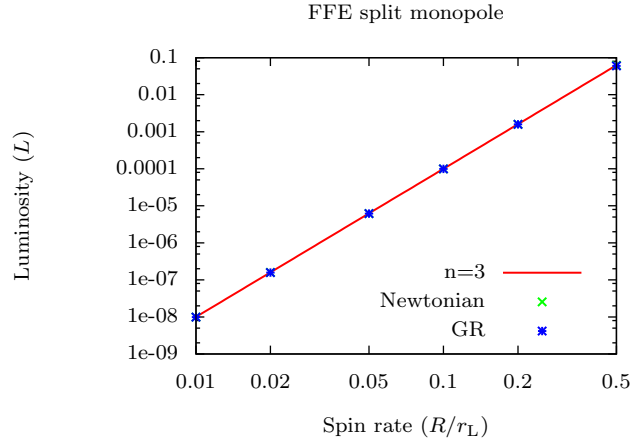}
\end{center}
\caption{Absolute Poynting flux of the force-free split monopole. The different approximations are marked by symbols, green for Newtonian and blue for GR and compared to the point dipole expectations $n=3$ in red.}
\label{fig:Indice_Split_FFE}
\end{figure}

\begin{figure}
\begin{center}
\input{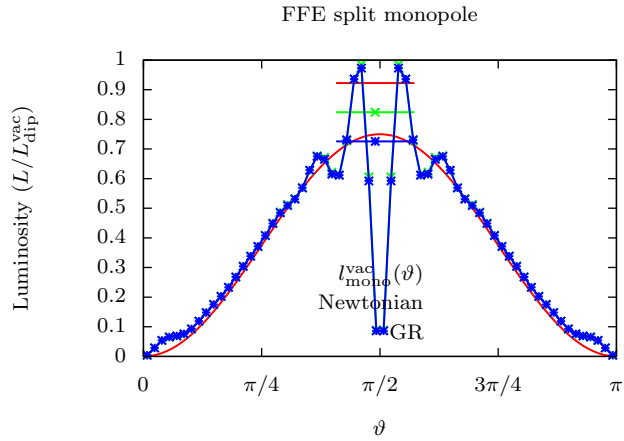}
\end{center}
\caption{Angular dependence of the Poynting flux for the force-free split monopole in flat space-time, green crosses, and in the slow rotation approximation, blue stars, with $R/\rlight=0.2$. The fit to $\sin^2\vartheta$ is shown in red.}
\label{fig:FFE_Split_Flux_Angulaire}
\end{figure}

\begin{figure}
 \centering
\input{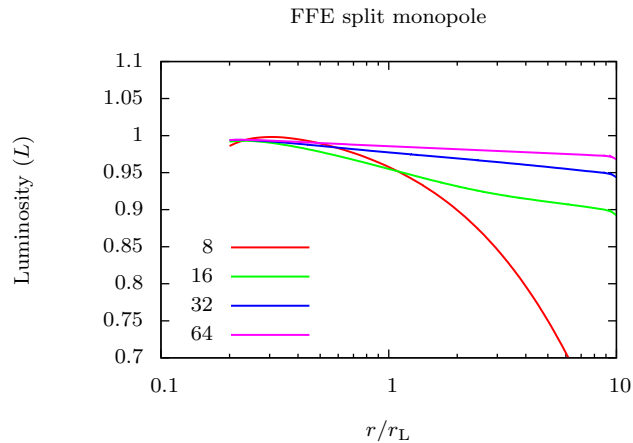}
 \caption{Luminosity of the force-free split monopole in the slow rotation approximation with $R/\rlight=0.2$ and for different numerical resolutions $N_\vartheta=\{8,16,32,64\}$ as given in the legend.}
 \label{fig:Luminosite_FFE_Split_Dissipation}
\end{figure}

The meridional magnetic field lines for the split monopole field are shown in fig.~\ref{fig:FFE_Split1}. Here again, as in the monopole case, the configuration in Boyer-Lindquist coordinates are not distinguishable. 
The associated Poynting flux should be the same as for the monopole case. This can be checked in fig.~\ref{fig:Luminosite_FFE_Split}. However we notice a decrease in the Poynting flux. This is due to the presence of the equatorial current sheet, forming a discontinuity that needs to be controlled by a filtering technique. This explains the dissipation of energy. 
The braking indexes for the Newtonian and general-relativistic split monopole solutions are compared and equal respectively to $n=3.001$ and $n=2.992$, obtained from fitting the absolute Poynting flux in fig.~\ref{fig:Indice_Split_FFE} to a power law in~$R/\rlight$.
The angular dependence of the Poynting flux for the force-free split monopole normalized to its total luminosity is equal to $l_{\rm mono}^{\rm vac}(\vartheta)$
and represented in fig.~\ref{fig:FFE_Split_Flux_Angulaire}. Because of the presence of a current sheet in the equatorial plane, the series expansion has trouble to adjust to such a discontinuity and the Gibbs phenomenon appears. This effect renders the convergence to the true solution very difficult as the error does only decrease linearly with the number of discretization points~$N_\vartheta$ in the colatitude.
In order to see the effect of the resolution on the accuracy of the luminosity, we looked at the convergence of the Poynting flux depending on the grid resolution in $\vartheta$ as shown in fig.~\ref{fig:Luminosite_FFE_Split_Dissipation}. Although the presence of the discontinuity, the Poynting flux converges to a constant value with only~$N_\vartheta=32$. The magnetic topology is not well represented along the equatorial plane but the global dynamics of the radiated power is still satisfactory.

\subsection{Dipole}

To finish this discussion, we summarize the new results about the rotating dipolar magnetosphere in the two limiting cases of an aligned and a perpendicular rotator. The main results of the simulations are the Poynting fluxes compared with respect to the Newtonian and general-relativistic geometry. Finally we will show an example of dependence with respect to the obliquity.

In order to emphasize the discrepancies between Newtonian gravity and general relativity, in this last section, we normalize the Poynting flux with respect to the aligned force-free rotator in flat spacetime given by
\begin{equation}
L_{\rm dip}^{\rm FFE} = \frac{3}{2} \, L_{\rm dip}^{\rm vac}
\end{equation} 
therefore better assessing the differences between both approximations.

\subsubsection{Aligned rotator}

\begin{figure}
\begin{center}
\input{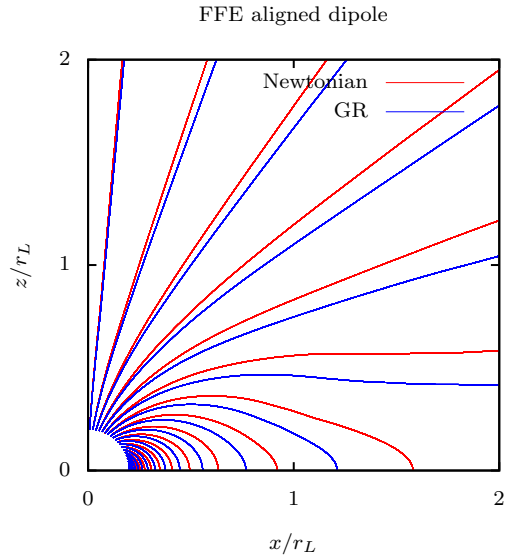} 
\end{center}
\caption{Meridional field lines for the force-free aligned dipole in flat space-time, red solid line, and in the slow rotation approximation, blue solid line, with $R/\rlight=0.2$.}
\label{fig:FFE_Dipole_Aligned1}
\end{figure}

\begin{figure}
\begin{center}
\input{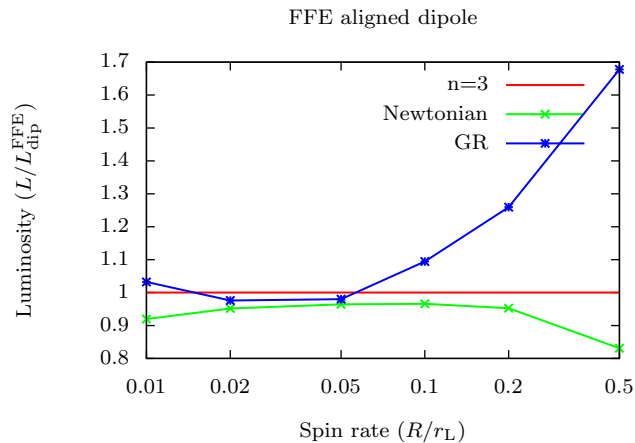}
\end{center}
\caption{Poynting flux of the force-free aligned rotator for $R/\rlight=\{0.01, 0.02, 0.05, 0.1, 0.2, 0.5\}$ and normalized to $L_{\rm dip}^{\rm FFE}$. The different approximations are shown in green for Newtonian and blue for GR and compared to the point dipole expectations $n=3$ in red.}
\label{fig:Poynting_Dipole_Align}
\end{figure}

\begin{figure}
\begin{center}
\input{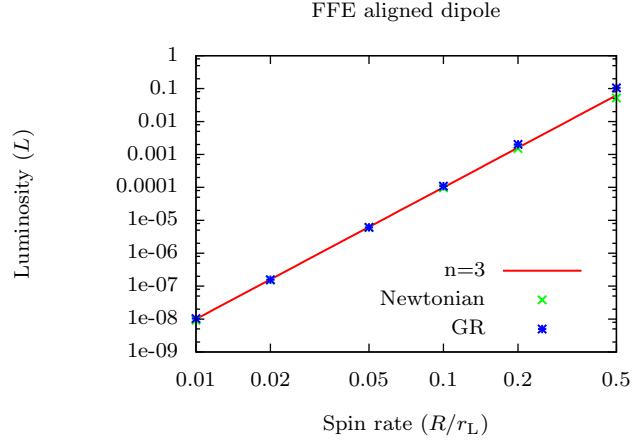}
\end{center}
\caption{Absolute Poynting flux of the force-free aligned dipole rotator. The different approximations are marked by symbols, green for Newtonian and blue for GR and compared to the point dipole expectations $n=3$ in red.}
\label{fig:Indice_Dipole_Aligne_FFE}
\end{figure}

\begin{figure}
\begin{center}
\input{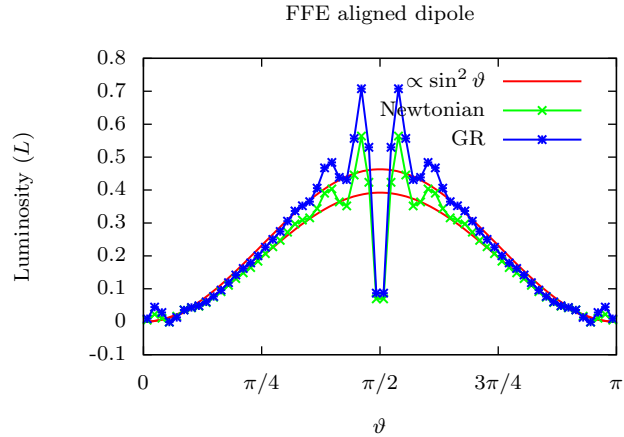}
\end{center}
\caption{Angular dependence of the Poynting flux for the FFE aligned dipole in flat space-time, green line, and in the slow rotation approximation, blue line, with $R/\rlight=0.2$. The fit to $\sin^2\vartheta$ is shown in red.}
\label{fig:FFE_Dipole_Aligned_Flux_Angulaire}
\end{figure}

\begin{figure}
 \centering
\input{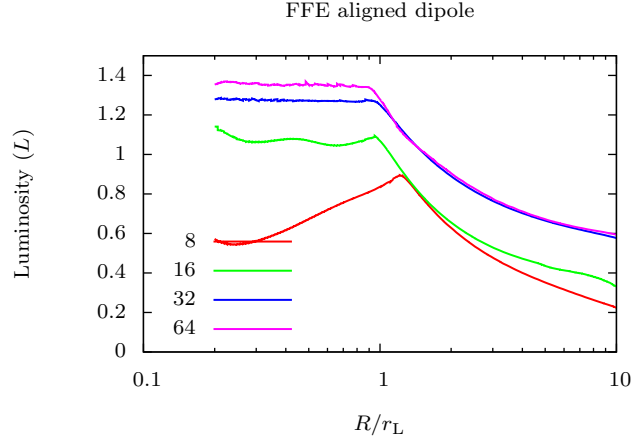}
 \caption{Luminosity of the force-free aligned dipole in the slow rotation approximation with $R/\rlight=0.2$ and for different numerical resolutions $N_\vartheta=\{8,16,32,64\}$ as given in the legend.}
 \label{fig:Luminosite_FFE_Dipole_Dissipation}
\end{figure}

The aligned rotator has been investigated by many authors in flat space-time. Here we present new simulations including general relativity. First we compare the magnetic field topology of the non relativistic and the general relativistic configurations. Examples of meridional field lines are shown in fig.~\ref{fig:FFE_Dipole_Aligned1} for flat space-time in solid red line and for curved space-time in solid blue line for a rotation rate of $R/\rlight=0.2$. General relativity leads to a compression of the field lines towards the neutron star as already observed in the vacuum case. Qualitatively, the topology does not significantly changes in both cases but the size of the polar cap and the curvature of the field lines are nevertheless affected. Note that because of the unavoidable numerical resistivity, some magnetic field lines still close outside the light cylinder.
The Poynting fluxes corresponding to the aligned rotator for several spin parameters $R/\rlight=\{0.01,0.02,0.05,0.1,0.2,0.5\}$ are described in fig.~\ref{fig:Poynting_Dipole_Align}. The current state of our code is pushed to its limit for very slow rotators with~$R/\rlight\lesssim0.02$ because the size of the polar caps become very small, requiring higher resolutions, thus smaller radial grids and much smaller time steps. The computation time becomes prohibitive and we were unable to perform high resolution simulations within a reasonable CPU time.
Nevertheless, let us have a look at the braking indexes for the Newtonian and general-relativistic aligned dipole. They are respectively $n=2.97$ and $n=3.12$, obtained from the fit shown in fig.~\ref{fig:Indice_Dipole_Aligne_FFE} for the absolute Poynting flux.
The angular dependence of the Poynting flux for the aligned dipole is depicted in fig.~\ref{fig:FFE_Dipole_Aligned_Flux_Angulaire}. As for the split monopole, the presence of a current sheet in the equatorial plane renders the convergence difficult in this region. Nevertheless, elsewhere the solution is satisfactory. The variation of the Poynting flux with the colatitude~$\vartheta$ has already been reported by \cite{2013MNRAS.435L...1T} for 3D magnetohydrodynamical (MHD) simulations. Moreover \cite{2015arXiv150301467T} looked at some analytical expressions summarizing the bunch of results obtained from MHD and force-free simulations of an oblique rotator. Our plots are in good qualitative agreements with their findings. Dissipation in the equatorial current sheet prevents us to see an appreciable deviation from the monopole angular pattern as for instance claimed by these authors or by \cite{2006MNRAS.368.1055T}.
In order to see the effect of the resolution on the accuracy of the luminosity, we looked at the convergence of the Poynting flux depending on the grid resolution in~$\vartheta$ as shown in fig.~\ref{fig:Luminosite_FFE_Dipole_Dissipation}. Dissipation is reduced when the number of points is increased as expected. A reasonable number of grid points $N_\vartheta=32$ or $N_\vartheta=64$ is sufficient to obtain good accuracy inside the light-cylinder. Nevertheless, outside we always get dissipation due to the presence of the current sheet in the equatorial plane. Our results seem a bit more dissipative than those obtained by the spectral code of \cite{2012MNRAS.423.1416P}. This is mostly due to the coarser grid at the outer edge (remind the geometric series increase in the radial size of the cells) and the lower number of discretization points in the colatitude.

\begin{figure}
\begin{center}
\input{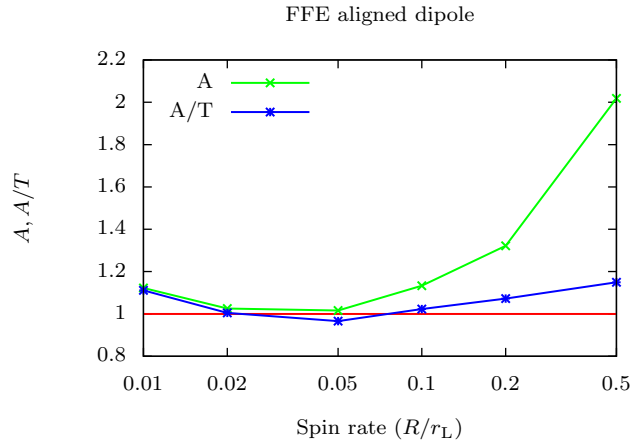}
\end{center}
\caption{Comparison of the increase in total luminosity as obtained by the simulation for the general relativistic case, solid green line, and as predicted by the increase in the transverse magnetic field, solid blue line. For reference the $L=1$ line is shown in red.}
\label{fig:Dipole_FFE_Aligne_Explication}
\end{figure}

The increase in total luminosity can in part be explained by the increase in the transverse magnetic field in the vicinity of the light-cylinder as was already discussed for the vacuum perpendicular rotator in the previous section. This is shown in fig.~\ref{fig:Dipole_FFE_Aligne_Explication} where in solid green curve we see the increase in luminosity induced by the curved spacetime and the possible explanation in solid blue line in terms of transverse field at the light-cylinder.

Comparing our results with \cite{PhysRevD.89.084045}, we found a small discrepancy between their enhancement of the spindown luminosity and our results. In their units, our compactness~$\Xi$ corresponds to their compaction~$C=\Xi/2=\Rs/(2\,R)=0.25$ and the rotation rate of the neutron star is~$\bar\Omega=R/\rlight$. In the general relativistic case, we always obtain spindown luminosities that are higher than the values they get. For $\bar\Omega=R/\rlight=0.1$ the enhancement is about 10\%-15\% in our simulations and in \cite{PhysRevD.89.084045} results but for larger spin rate, for instance $\bar\Omega=R/\rlight=0.2$ we get about 30\% whereas they got 17\%. The deviation is even more pronounced for spin rates $\bar\Omega=R/\rlight=0.5$ (which are however unrealistic according to observations). However as these authors claimed, they expect ``stiffer equations of state and more rapidly rotating neutron stars lead to even larger enhancements''. Therefore the exact spindown value depends on the equation of state used for the polytrope. In our case, we did not solve for the electromagnetic field inside the star so we must interpret the difference in enhancement by an effect from the boundary conditions imposed on the stellar surface. We fixed them to the exact general relativistic non rotating dipole for the inside solution whereas they computed it from a prescription for the stellar structure. In our case, the slow rotation approximation becomes questionable for $R/\rlight=0.5$. Indeed, at this spin rate the ratio between kinetic energy~$T=I\,\Omega^2/2$ and gravitational binding energy~$W=-3\,G\,M^2/(5\,R)$ for a homogeneous sphere of radius~$R$ given by
\begin{equation}
\frac{T}{|W|} = \frac{2}{3} \, \frac{R}{\Rs} \, \left( \frac{R}{\rlight} \right)^2
\end{equation} 
is about $0.33$, far from the $T \ll |W|$ condition. Thus rotational effect cannot be ignored for an accurate description of the neutron star shape and the results we found should not be taken to precisely.

\subsubsection{Orthogonal rotator}

\begin{figure}
\begin{center}
\input{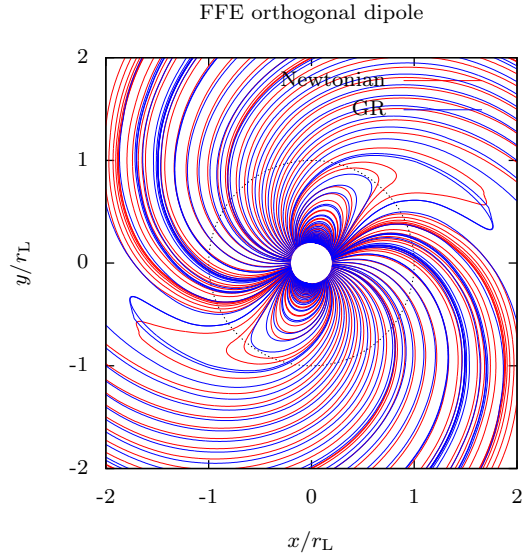} 
\end{center}
\caption{Equatorial field lines for the force-free perpendicular dipole in flat space-time, red solid line, and in the slow rotation approximation, blue solid line, with $R/\rlight=0.2$.}
\label{fig:FFE_Dipole_Perp2}
\end{figure}

\begin{figure}
\begin{center}
\input{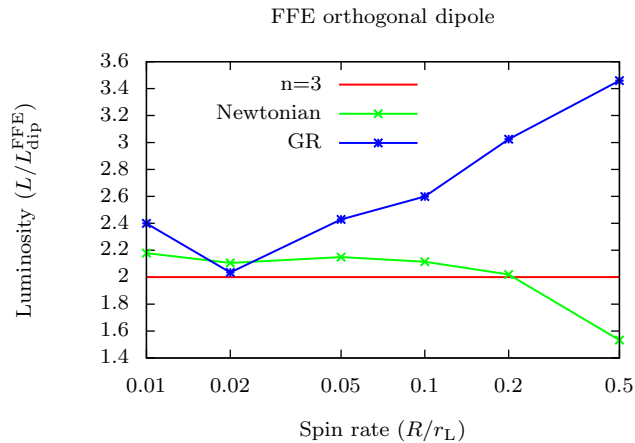}
\end{center}
\caption{Poynting flux of the force-free orthogonal rotator normalized to $L_{\rm dip}^{\rm FFE}$ and for $R/\rlight=\{0.01,0.02,0.05,0.1,0.2,0.5\}$. The different approximations are marked by symbols, green for Newtonian and blue for GR, and compared to the point dipole expectations $n=3$, red line.}
\label{fig:Poynting_Dipole_Perp}
\end{figure}

\begin{figure}
\begin{center}
\input{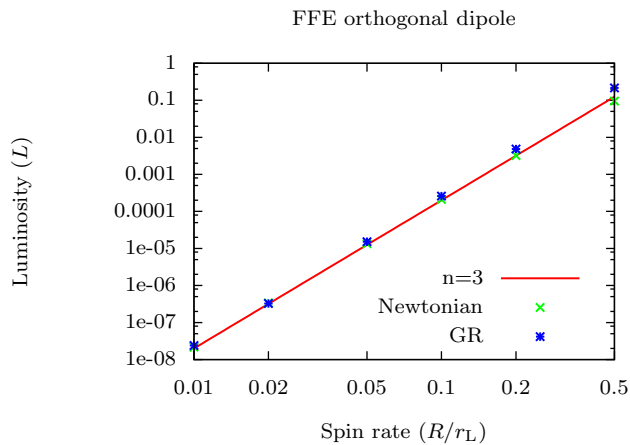}
\end{center}
\caption{Absolute Poynting flux of the force-free orthogonal rotator. The different approximations are marked by symbols, green for Newtonian and blue for GR. They are compared to the point dipole expectations $n=3$, red line.}
\label{fig:Indice_Dipole_Perp_FFE}
\end{figure}

Investigations of the orthogonal rotator follow the same line as the aligned case. Examples of equatorial field lines are shown in fig.~\ref{fig:FFE_Dipole_Perp2} for flat space-time in red and for curved space-time in blue, for a rotation rate of $R/\rlight=0.2$.
The Poynting fluxes corresponding to the perpendicular rotator for several spin parameters $R/\rlight=\{0.01,0.02,0.05,0.1,0.2,0.5\}$ are described in fig.~\ref{fig:Poynting_Dipole_Perp}.
The braking indexes for the Newtonian and general-relativistic perpendicular dipole are respectively $n=3.04$ and $n=3.19$, obtained from the fits shown in fig.~\ref{fig:Indice_Dipole_Perp_FFE}.

\begin{figure}
\begin{center}
\input{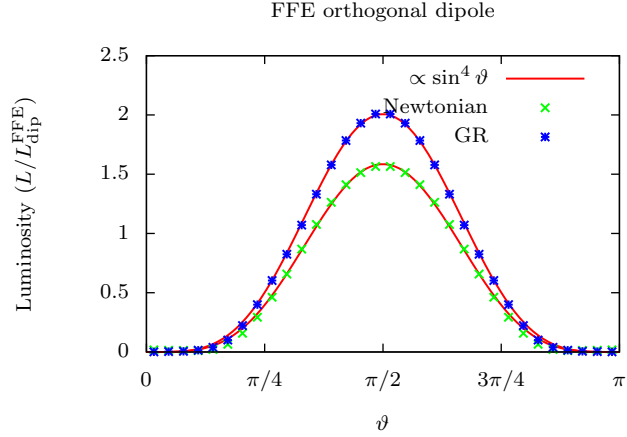}
\end{center}
\caption{Angular dependence of the Poynting flux for the FFE orthogonal dipole in flat space-time, in green, and in the slow rotation approximation, in blue, with $R/\rlight=0.2$. A fit to $\sin^4\vartheta$ is proposed in red.}
\label{fig:FFE_Dipole_Perp_Flux_Angulaire}
\end{figure}

\begin{figure}
\begin{center}
\input{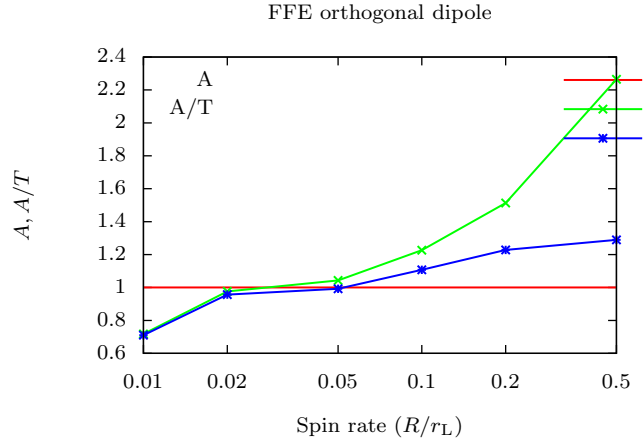}
\end{center}
\caption{Comparison of the increase in total luminosity as obtained by the simulation, green line, and as predicted by the increase in the transverse magnetic field, blue line. For reference the $L=1$ line is shown in red.}
\label{fig:Dipole_Perp_FFE_Explication}
\end{figure}

The angular dependence of the Poynting flux is depicted in fig.~\ref{fig:FFE_Dipole_Perp_Flux_Angulaire}. Contrary to the split monopole or aligned dipole, the current sheet in the equatorial plane has disappeared thus improving the convergence in the whole computational domain. We found that the flux behaves like $L(\vartheta) \propto \sin^4\vartheta$ to good accuracy. Such simple fit to the luminosity was also proposed by \cite{2013MNRAS.435L...1T}.
The increase in total luminosity can in part be explained by the increase in the  transverse magnetic field in the vicinity of the light-cylinder as shown in fig.~\ref{fig:Dipole_Perp_FFE_Explication}.

\subsubsection{Oblique rotator}

\begin{figure}
\begin{center}
\input{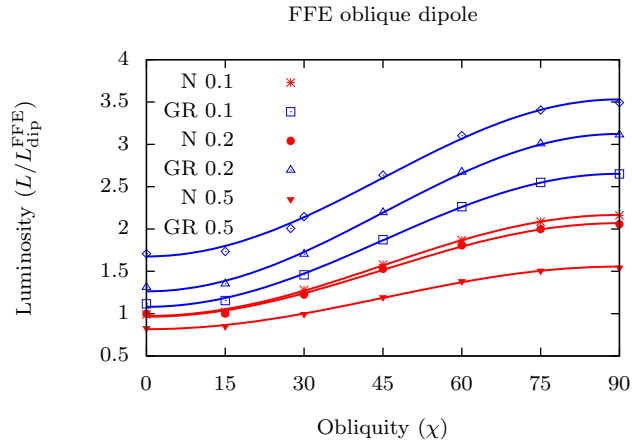}
\end{center}
\caption{Poynting flux of the force-free oblique rotator for different obliquities~$\chi$ and normalized to $L_{\rm dip}^{\rm FFE}$. Points are taken from the simulations and the solid lines are the best fits with $R/\rlight=\{0.1, 0.2, 0.5\}$. Red corresponds to Newtonian approximation (N) and blue to general relativity (GR).}
\label{fig:Poynting_Dipole_Oblique_FFE_R1}
\end{figure}

To conclude, we performed a set of simulations for oblique rotators in the case $R/\rlight=\{0.1, 0.2, 0.5\}$ and for obliquities~$\chi=\{0\degr,15\degr,30\degr,45\degr,60\degr,75\degr,90\degr\}$ to check the dependence of the spin-down luminosity on the geometry in the force-free limit. The results are summarized in fig.~\ref{fig:Poynting_Dipole_Oblique_FFE_R1}. All the curves can be fitted with an expression like
\begin{equation}
\frac{L}{L_{\rm dip}^{\rm FFE}} = a + b \, \sin^2\chi
\end{equation}
where $a$ and $b$ are constants. This behaviour is reminiscent of the vacuum dipole rotator in flat space-time. Best fits parameters for $a$ and $b$ are given in table~\ref{tab:FitFluxFFE}. As a general trend, curvature of space-time increases the overall Poynting flux.
The coefficient~$a$ is not constant and approximately equal to unity as claimed in the simulations done by \cite{2014MNRAS.441.1879P}. We found a small decrease with respect to $R/\rlight$ consistent with the vacuum oblique rotator.
\begin{table}
\centering
\begin{center}
\begin{tabular}{ccc}
\hline
$R/\rlight$ & Newtonian & GR \\
\hline
\hline
0.1 & 0.972 / 1.193 & 1.080 / 1.570 \\
0.2 & 0.847 / 1.184 & 0.946 / 2.094 \\
0.5 & 0.818 / 0.737 & 1.650 / 1.888 \\
\hline
\end{tabular}
\end{center}
\caption{Best fit parameters~$a/b$ for the Poynting flux $L(\chi)/L_{\rm dip}^{\rm FFE} = a + b\,\sin^2\chi$ of the FFE dipole rotator in Newtonian and general relativistic (GR) cases.}
\label{tab:FitFluxFFE}
\end{table}

\section{Conclusions}
\label{sec:Conclusion}

General-relativistic effects are important to understand the electrodynamical processes in the vicinity of neutron stars. This was already known in the case of a vacuum rotator for which the Poynting flux increases monotonically with increasing curvature and frame-dragging effects. These conclusions remain true for a force-free pulsar magnetosphere. The increase in spin-down luminosity can reach a factor up to 2 depending on the neutron star period, through the ratio~$R/\rlight$. The overall effect will be a change in the estimate of the surface dipolar magnetic field by several tenth of percents compared to flat space-time expectations. For the oblique rotator we retrieve the law $L(\chi)/L_{\rm dip}^{\rm FFE} = a+b\,\sin^2\chi$ for the Poynting flux, reminiscent of the Newtonian vacuum dipole except for slight changes in the constants $a$ and $b$ ($a$ vanishing in vacuum). We were also able to investigate slow rotators with ratio~$R/\rlight$ as small as $0.01$. This helped us to compute the braking index~$n$ in vacuum and in the force-free limit. We found values always close to the fiducial braking index of $n=3$.

Although force-free magnetospheres represent a useful first step to describe the global structure of relativistic plasmas evolving in the electromagnetic field of neutron stars, they do not furnish any information about the sites where particle acceleration and therefore radiation is supposed to occur. A good compromise between full PIC or MHD calculations and our simple force-free model, taking into account plasma inertia and finite temperature, would be to used a resistive prescription for the current density as already done by several authors. This extension is left for future work.

\section*{Acknowledgements}

I am grateful to the referee for helpful comments. This work has been supported by the French National Research Agency (ANR) through the grant No. ANR-13-JS05-0003-01 (project EMPERE). It also benefited from the computational facilities available at Equip@Meso of the Universit\'e de Strasbourg.


\label{lastpage}

\end{document}